\documentclass[aps,prb,onecolumn,showpacs,showkeys,floatfix,groupedaddress]{revtex4-2}

\usepackage{amsfonts,fancybox,pgf}
\usepackage{amssymb}
\usepackage{float}
\usepackage{footnote}
	\usepackage{amsmath}
	\usepackage{makeidx}
	\usepackage{amsfonts}
	\usepackage[ansinew]{inputenc}
	\usepackage[usenames,dvipsnames]{pstricks}
	\usepackage{subfigure}
	\usepackage{graphicx}
	\usepackage{here}
    \usepackage{epstopdf}
	\usepackage{epsfig}
	\usepackage{pst-grad} 
	\usepackage{pst-plot} 
	\usepackage[colorlinks,hyperindex]{hyperref}
	\usepackage[english]{babel}

	\hypersetup
		{
		colorlinks,%
		citecolor=blue,%
		linkcolor=black,%
		urlcolor=black,%
	}

\makeindex

\begin{document}

\title{\normalsize A Renormalized Ginzburg-Landau Framework for Dimensional Crossover and Fluctuation Specific Heat in High-$T_c$ Superconductors in a Magnetic Field}

\author{R. M. Keumo Tsiaze$^{a,b, c}$, J. E. Danga$^{c, d}$, L.C. Fai$^{c, d}$}

\address{$^{a}$ International Chair in Mathematical Physics and Applications, (ICMPA-UNESCO Chair), University of Abomey-Calavi, 072 P O Box 50, Cotonou, Republic of Benin}
\address{$^{b}$ Laboratory of Mechanics, Materials and Structures, Faculty of Science, University of Yaound\'{e} I, P.O. Box 812, Yaound\'{e}, Cameroon}
\address{$^{c}$ Quantum Materials and Computing Group - QMaCG, P.O. Box 70 Bambili, Northwest Region, Cameroon.}
\address{$^{d}$  Laboratory of Condensed Matter-Electronics and Signal Processing (LAMACET), Department of Physics, Faculty of Science, University of Dschang, P.O. Box 67 Dschang, Cameroon. \\}
\email{keumoroger@gmail.com} 

\begin{abstract}
This paper presents a theoretical analysis of phase transitions and critical phenomena in high-temperature superconductors using a renormalized Ginzburg-Landau framework. Rather than assuming a conventional linear temperature dependence, we treat the quadratic coefficient as a self-consistent, Hartree-renormalized quantity determined by fluctuation-loop corrections. This renormalization regularizes mean-field divergences and yields a finite, dimensionality-dependent specific-heat anomaly near the transition. When an external magnetic field is applied, minimal coupling quantizes order-parameter fluctuations into discrete Landau levels, reducing the effective dimensionality via an effective spectral dimension. In this framework, the intrinsic fluctuation-coupling strength self-consistently determines the temperature width of the critical Ginzburg region, while the cyclotron energy of the quantized fluctuations uniquely establishes both the specific-heat peak position and the upper-critical-field crossover boundary. Comparison with $\text{YBa}_2\text{Cu}_3\text{O}_{7-\delta}$ experimental data demonstrates that this approach captures the suppression of the sharp mean-field discontinuity and accurately reproduces the vortex-lattice topological structure under finite magnetic fields.

\end{abstract}

\vspace{0.5pc}


%
\maketitle
%

\section{Introduction}

The miniaturization of devices has enabled significant scientific discoveries and the development of novel functionalities that emerge from the microscopic properties of materials. Reducing the thickness of a bulk material in one direction increases its effective density in the remaining dimensions, thereby facilitating further dimensional reduction. Consequently, the investigation and design of materials with diverse geometries and symmetries, which frequently yield new functionalities and notable properties, particularly regarding phase transitions, remains a central focus of research \cite{Landau1, HohenbergPC, Landau2, Ginzburg, Stanley, Privman, Larkin, Hohenberg1, Wu}. 

Significant size effects are inherent to low-dimensional systems even without an external field. In contrast, bulk materials exhibit comparably strong size effects only when a magnetic field is applied, as Landau quantization collapses the degrees of freedom transverse to the field, driving the system toward an effectively lower-dimensional fluctuation regime. Recent studies have renormalized the corresponding scaling amplitude by incorporating order-parameter fluctuations and particle scattering through a material-specific parameter $\eta$ \cite{Keumo3}. This renormalization alters the prefactor that governs the fluctuation-specific heat, while the underlying critical exponent is determined by the system's effective dimensionality, in accordance with standard scaling theory. The shapes of the resulting specific-heat anomalies differ significantly
between dimensionalities: zero- and one-dimensional systems display properties characteristic of discretized, finite energy-level spectra, whereas two-, three-, and four-dimensional systems retain the qualitative structure of their respective mean-field fixed points, albeit with quantitatively and, near $D=4$, qualitatively modified critical behavior.

Applying a magnetic field to a superconductor modifies its thermodynamic properties and alters the fluctuation-specific heat by shifting the phase boundary, affecting vortex dynamics, and suppressing Cooper-pair fluctuations above the transition. A sufficiently strong field restricts the degrees of freedom available to the order parameter, resulting in a nominally three-dimensional (3D) system exhibiting effectively one-dimensional critical dynamics near the transition. Without a field, Ginzburg-Landau mean-field theory predicts that superconducting fluctuations diverge as $(T/T_{c_0}-1)^{-1/2}$, the standard 3D Gaussian exponent. The introduction of an external field shifts the transition temperature to $T_{c_B}$ and confines thermal fluctuations along the field direction. This quasi-one-dimensional geometry leads to a much sharper divergence, scaling
as $(T/T_{c_B}-1)^{-3/2}$, consistent with the field-induced dimensional reduction discussed in Sec.~\ref{sec:theory} and a well-documented history \cite{Thouless, Quader, Gupta, Shenoy}. In this review, we revisit this phenomenon by treating the Ginzburg--Landau coefficients as functions of the system's intrinsic characteristics, enabling precise calculation of the fluctuation-induced specific heat while accounting for dimensional effects, material-dependent energy scales, and non-monotonic anomalies.

In conventional mean-field theory, the quadratic coefficient is assumed to depend linearly on temperature, $r_0(T-T_{c_0})$, an assumption that leads to unphysical divergences in the specific heat near $T_{c_0}$. The renormalized coefficient $r(D,T)$, as developed in Ref.~\cite{Keumo3} and applied in the presence of a magnetic field, incorporates strong fluctuation corrections and regularizes the divergence for all physically relevant dimensionalities $D<4$. At $D=4$, the genuine upper critical dimension, the renormalized coefficient develops a pole, delineating the boundary beyond which mean-field theory becomes exact. In the presence of a magnetic field, discrete Landau levels further modify the density of states, producing oscillations in the specific heat that are markedly more pronounced in two-dimensional (2D) systems, where the fluctuation sum retains its discrete Landau-level structure, than in 3D systems, where an additional integration over the field-parallel momentum smooths these oscillations.

A notable feature of unconventional systems, such as high-temperature superconductors, is the pronounced broadening of their transition in a magnetic field, which is interpreted here as arising from thermodynamic fluctuations near the transition. Within the present orbital-coupling framework, the applied field serves as a crossover parameter that reduces the effective dimensionality of the fluctuations, leading to discrete Landau-level structure in the specific heat. A fully quantitative treatment
of the accompanying spin (Zeeman) pair-breaking mechanism and of momentum-space pairing anisotropy specific to unconventional superconductors is beyond the scope of the current orbital formalism and is reserved for future work. For a field $B$ applied perpendicular to the superconducting planes, the electronic kinetic energy is quantized into
discrete Landau levels, $E_N=\hbar\omega_c(N+\tfrac12)$, each with degeneracy per unit area $e^*B/2\pi\hbar$. In the lowest-Landau-level approximation, valid for sufficiently strong fields (large $\omega_c=e^*B/m^*$) such that only $N=0$ is populated, the electronic structure is significantly altered. Here, $m^*=2m_e$ and $e^*=\pm2e$ denote
the mass and charge of a superconducting charge carrier (Cooper pair), with the sign of $e^*$ determined by whether the pair consists of electrons (negative) or holes (positive). In the analysis that follows, $e^*=-2e$ is used, corresponding to electron-like carriers.

This review develops a theoretical model of low-dimensional superconducting fluctuations in the presence of an external magnetic field. Rather than focusing on systems that are geometrically confined to one or zero dimensions, such as nanowires or quantum dots, it is demonstrated that a magnetic field applied to a nominally 3D bulk superconductor or to a 2D film drives the fluctuation problem to an effectively reduced dimensionality through Landau quantization of the degrees of freedom transverse to the field. For a bulk sample in a field, the transverse motion is fully quantized while the field-parallel direction remains free, resulting in effectively one-dimensional (1D) fluctuations. This effect rounds the specific-heat jump characteristic of ordinary 3D superconductors into a broadened, enhanced heat-capacity tail that extends well above the transition. For a thin film with the field applied perpendicular to the plane, both in-plane directions are quantized into Landau orbits, confining the fluctuating Cooper pairs to effective zero-dimensional (0D) droplets of radius $\xi(T)$.

The structure of this review is as follows. Section~\ref{sec:theory} introduces the field-modified Ginzburg--Landau theory, emphasizing the dimensional character of order-parameter fluctuations governed by Landau quantization, and derives the corresponding fluctuation-specific heat as an explicit function of dimensionality, temperature, and field. Representative parameter values are used to illustrate how dimensionality and thermal fluctuations jointly determine the thermodynamic response and how the
model captures physical regimes inaccessible to conventional mean-field theory. Section~\ref{sec:numerics} plots the fluctuation specific heat as a function of the fluctuation coupling $\eta$, the applied magnetic field $B$, and temperature $T$. Section~\ref{sec:experiment} compares these theoretical results with experimental specific-heat data for
YBa$_2$Cu$_3$O$_{7-\delta}$. Section~\ref{sec:conclusion} concludes the review.

\section{The renormalized Ginzburg--Landau theory: role of the $\varphi^4$ term}
\label{sec:theory}

The primary cause of standard Ginzburg--Landau theory's (GLT) failure is its inability to adequately describe systems with reduced dimensionality. For example, the standard GLT does not adequately explain the order of phase transitions in certain layered bulk materials \cite{Cybart}. Furthermore, it is insufficient for describing 1D
\cite{Scalapino} and quasi-2D systems, where novel states with unexpected properties emerge at non-zero temperatures \cite{Tinkham}. Another significant limitation is its inability to account for the enhancement of the specific-heat jump in YBa$_2$Cu$_3$O$_{7-\delta}$ (YBCO) \cite{Loram}, or the absence or disappearance of this jump in materials \cite{Keumo3} such as Ba$_{0.2}$K$_{0.8}$Fe$_2$As$_2$ \cite{Tanaka} and in certain families of high-$T_c$ superconductors, including TlBaCaCuO and BiSrCaCuO
\cite{Meingast}. The effects of critical fluctuations, their dynamics, and their influence on the physical properties of materials have been investigated in numerous studies \cite{Anatoly, Doniach, Zinn-Justin, Ma, Amit, Papon, Varlamov, Kleinert}. Nevertheless, challenges persist in universally assigning the parameters of the standard GLT to characteristic quantities, such as temperatures \cite{Poole}. The renormalization process improves the standard theory by introducing a correction term to the quadratic
coefficient, which is not included in the original model.

\subsection{Absence of an external magnetic field}

Within the Ginzburg--Landau framework, the thermodynamic properties of high-$T_c$ superconductors near $T_{c_0}$, including the specific heat, are described by an effective Hamiltonian formulated in terms of the complex superconducting order parameter $\varphi(\mathbf r)$. In the absence of an external magnetic field, the standard anisotropic-mass Ginzburg-Landau Hamiltonian $\mathcal H_{\mathrm{GL}}[\varphi]$ for a layered cuprate such as YBCO is given by the following spatial integral (see \cite{Keumo3} and other references therein):
\begin{equation}
\mathcal H_{\mathrm{GL}}[\varphi]=\mathcal H_n+\int d^D r\left[
\frac12 r(T)|\varphi|^2+\frac b4|\varphi|^4
+\frac{\hbar^2}{2m_x}\left(\frac{\partial\varphi}{\partial x}\right)^2
+\frac{\hbar^2}{2m_y}\left(\frac{\partial\varphi}{\partial y}\right)^2
+\frac{\hbar^2}{2m_z}\left(\frac{\partial\varphi}{\partial z}\right)^2
\right].
\label{eq:HGL0}
\end{equation}
This study does not present a fully microscopic treatment of high-$T_c$ superconductors, which generally exhibit a superconducting order parameter (OP) with $d$-wave pairing symmetry ($d_{x^{2}-y^{2}}$) and require an explicit momentum-space form factor or a multi-component OP. Instead, the anisotropic-mass Hamiltonian in Eq. (1) addresses the distinct crystallographic anisotropy in real space resulting from YBCO's layered structure, specifically the contrast between the CuO$_2$ planes and the $c$-axis, rather than the internal $d$-wave symmetry of the superconducting gap. The described theoretical framework, applicable near $T_{c_0}$ and in the long-wavelength, coarse-grained limit, remains the standard for modeling such properties.

The $\varphi^4$ potential [$V(\varphi)\propto \tfrac b4\varphi^4$] is crucial in studying spontaneous symmetry breaking, where the field develops a non-zero vacuum expectation value, playing a key role in models such as the Higgs mechanism. Solving the $\varphi^4$ term in quantum field theory involves handling a non-linear interaction that prevents an exact, closed-form solution. Instead, the term is treated using approximation techniques, primarily perturbation theory, to calculate scattering amplitudes, vacuum expectation values, and renormalization effects. Recent developments employing a decomposition method related to perturbative renormalization suggest that \cite{Keumo1, Keumo2}
\begin{equation}
|\varphi(q)|^4\approx 6\langle|\varphi(q)|^2\rangle|\varphi(q)|^2,
\label{eq:phi4}
\end{equation}
under the assumption that Fourier components interact solely through the mean field generated by other modes. The factor of 6 accounts for all possible contractions that result from the intrinsic properties of real order parameters. 

When the $\varphi^4$ term is associated with redundant fluctuations, the quartic term represents an interaction among the Fourier components of the OP. The $|\varphi(q)|^4$ term represents a free-energy contribution from fluctuation interactions. Therefore, the functional integral in Eq.~\eqref{eq:HGL0} is evaluated over a Gaussian field:
\begin{equation}
r(T)\rightarrow r(T)+3b\langle|\varphi|^2\rangle,
\label{eq:hartree}
\end{equation}
enabling analytical computation.  New quadratic coefficients are obtained by equating the quadratic terms in Eq.~\eqref{eq:HGL0} with $|\varphi(q)|^4$ evaluated using a Hartree-type approximation, where the modulus brackets represent an average computed via functional integration over all possible fluctuations in the OP. After straightforward calculations, the following estimates for $r(D,T)$ in the dirty limit and in low dimensions, in the absence of a field, are obtained. The quantity $\langle|\varphi|^2\rangle$ exhibits a strong dependence on the dimensionality of the system \cite{Keumo3}. This term partially characterizes the interaction between the in-plane and out-of-plane components of the OP in cuprates, as discussed below. The expectation value $\langle|\varphi|^2\rangle$ is calculated self-consistently, leading to
the self-consistent equation
\begin{equation}
r(D,T)-\eta\left(\frac{T}{T_{c_0}}\right)
\left[\frac{r(D,T)}{r_0T_{c_0}}\right]^{D/2-1}-r_0(T-T_{c_0})=0.
\label{eq:selfconsistent}
\end{equation}
with solutions
\begin{equation}
r(D,T)=
\begin{cases}
\dfrac12\Big[r_0(T-T_{c_0})\pm\sqrt{r_0^2(T-T_{c_0})^2+4\eta r_0T}\Big], & D=0,\\[2mm]
r_0T_{c_0}\left[\dfrac{f(T)}{6r_0T_{c_0}}+\dfrac{2r_0(T-T_{c_0})}{f(T)}\right]^2, & D=1,\\[2mm]
\eta\left(\dfrac{T}{T_{c_0}}\right)+r_0(T-T_{c_0}), & D=2,\\[2mm]
\dfrac{\eta T\pm\sqrt{\eta^2T^2+4r_0^2T_{c_0}^3(T-T_{c_0})}}{2r_0T_{c_0}^2}+r_0(T-T_{c_0}), & D=3,\\[2mm]
r_0\left[1-\dfrac{\eta T}{r_0T_{c_0}^2}\right]^{-1}(T-T_{c_0}), & D=4.
\end{cases}
\label{eq:rDT}
\end{equation}
Eq.~\eqref{eq:selfconsistent} enables modeling of the system as semi-homogeneous, thereby simplifying complex mathematical and physical representations by averaging non-uniform properties, such as diffusion rates or state characteristics, across defined domains or dimensions. Microscopic derivations relevant to systems exhibiting pronounced
fluctuation regimes yield these renormalized quadratic coefficients. These modifications account for both the system's dimensionality and the scale of fluctuations. The renormalized quadratic coefficient, as determined by Eq.~\eqref{eq:rDT}, governs the system's overall behavior and produces a real transition temperature when anisotropic effects are included. For the case $D=1$, the function $f(T)$ is given by
\begin{equation}
f(T)=r_0^2T_{c_0}\left[108\eta T+12\sqrt{81\eta^2T^2-12r_0^2T_{c_0}(T-T_{c_0})^3}\right]^{1/3}.
\end{equation}
This expression facilitates the analysis of the behavior of $r(1,T)$ in numerical studies.

The dimensionality-dependent renormalization of the quadratic coefficient, $r(D,T)$, possesses physical significance that extends beyond a formal generalization of the mean-field expression $r(T)=r_0(T-T_{c_0})$. In the standard Ginzburg--Landau framework, $r(T)$ results from a simple expansion of the free energy and does not capture how order-parameter fluctuations propagate spatially. Retaining $D$ explicitly therefore allows the strength of fluctuation renormalization to be quantified directly as a
function of dimensionality, rather than being fixed a priori at $D=3$. This approach is not solely a mathematical refinement; solutions for $D=1$--$4$ exhibit qualitatively distinct behavior, culminating in the divergence of $r(4,T)$ at a finite temperature. This divergence directly indicates that $D=4$ is the upper critical dimension, above which
mean-field theory becomes exact, and fluctuation corrections are no longer relevant in the renormalization-group framework. The model introduces a second critical dimension, $d_c = 2$, below which fluctuations become highly significant. In contrast, at $d_U = 4$, fluctuations are less influential and require only consideration. In many systems, $d_c = 2$ marks the point below which fluctuations are sufficiently strong to eliminate long-range order. The Mermin-Wagner-Hohenberg theorem \cite{Mermin, Hohenberg} establishes that for $d \leq 2$, continuous symmetries cannot be spontaneously broken at finite temperature due to strong fluctuations \cite{Patashinski}. This principle is essential for delineating the limitations of mean-field assumptions and for determining when more advanced analytical methods are necessary.  When the quadratic coefficient in the standard GLT is modeled as a linear function of temperature, the resulting derivative remains constant.

 Within the renormalized formalism, $r'(D,T)\equiv dr(D,T)/dT$ depends on both temperature and the dimensionality of the system. When the derivative becomes temperature-dependent, the thermodynamic landscape becomes nonlinear, potentially indicating non-mean-field critical behavior and the presence of competing orders near the transition. According to Eq.~\eqref{eq:rDT}, the derivative of the quadratic coefficient functions as an effective scaling parameter, termed a ``pseudo-temperature.'' Defining this coefficient derivative as a thermal variable facilitates the analysis of system complexity across different states and helps identify scaling laws. This approach enables models to assess the influence of thermal fluctuations and energy parameters on system stability.

This framework is particularly applicable to layered cuprates such as YBCO synthesized as YBa$_2$Cu$_3$O$_{7-\delta}$ with $0\le\delta\le0.18$, whose superconductivity is not strictly 3D. Near $T_{c_0}$, when the coherence length along the $c$-axis is shorter than the interlayer spacing $d$, the CuO$_2$ planes function as effectively decoupled 2D sheets. At temperatures further from $T_{c_0}$, interlayer coupling restores 3D behavior. A dimensionality-dependent $r(D,T)$ provides an analytically controlled method for describing the 2D-to-3D crossover, as an alternative to a fully anisotropic 3D treatment. This approach also aligns with the Ginzburg criterion, which determines the width of the fluctuation-dominated region around the critical point and is strongly dependent on $D$ near the upper critical dimension. The combination of YBCO's short coherence length
and reduced effective dimensionality makes this fluctuation region experimentally accessible, in contrast to conventional BCS superconductors, where the analogous region is extremely narrow.

In the absence of an applied magnetic field, the superconducting phase transition in 2D systems is characterized by the unbinding of thermally generated vortices, a phenomenon known as the Kosterlitz--Thouless (KT) phase transition. In 3D systems, topological defects manifest as extended line defects rather than point defects, and standard thermal fluctuations or microscopic interlayer couplings result in conventional 3D long-range ordering, typically described by the 3D-XY or Heisenberg universality classes. Further details will be provided in subsequent sections.

The progression from $r_0(T-T_{c_0})$ to $r(D,T)$ given by Eq.~\eqref{eq:rDT} establishes a mathematical framework for renormalizing the quadratic coefficient in Landau theory in its classical form. This approach reconciles classical mean-field approximations with strong, dimensionally dependent thermal fluctuations. The resulting framework supports both a priori (predictive) and a posteriori (corrective) fluctuation-correction estimates. $r(D,T)$ is specifically adjusted to incorporate the system's configuration, including dimensionality and anisotropy. It acts as a self-consistent condition in nonlinear polynomial equations, defining the behavior of the OP and physical properties such as specific heat and susceptibility near phase transitions. This approach often finds that renormalized coefficients in 0D- and 1D systems are consistently positive regardless of temperature, demonstrating how the system avoids the instability predicted by standard mean-field theory. The calculation of fluctuation-specific heat within the Ginzburg-Landau framework requires integration over Gaussian fluctuations of the OP. When a magnetic field is applied, spatial quantization into Landau levels restricts the allowed states, thereby modifying the fluctuation-specific heat from the zero-field 3D power-law or 2D logarithmic behavior to a field-dependent form.

\subsection{Presence of an external magnetic field}

In condensed matter physics, the quartic $\varphi^4$ interaction serves as the canonical model for spontaneous symmetry breaking of a complex order parameter. Because $\varphi$ is complex, the symmetry is $U(1)$ rather than $\mathbb Z_2$. An external magnetic field couples to $\varphi$ not by aligning it, as in discrete-symmetry systems, but through minimal coupling of the gradient term to the vector potential $\mathbf A$, along with a Zeeman coupling to the electron spin. The orbital (minimal-coupling) contribution modifies the effective mass and quantizes the orbital motion perpendicular to $\mathbf B$ into discrete Landau levels, while the Zeeman term directly breaks time-reversal symmetry. The following analysis retains only the orbital coupling, consistent with the neglect of spin-paramagnetic pair breaking, as discussed later in this section. A
fully quantitative treatment of that mechanism would require an explicit Zeeman term, which is reserved for future work. Applying the transformation
\begin{equation}
r(T)|\varphi|^2\rightarrow r(D,T)|\varphi^*|^2,
\label{eq:transform}
\end{equation}
the general form of the free-energy density is preserved, with coefficients reparametrized by dimensionality as previously discussed. In the presence of a magnetic field, orbital quantization eliminates the continuous momentum integral over the plane(s) perpendicular to $\mathbf B$ from the fluctuation problem. Consequently, the coefficient entering the fluctuation-specific heat becomes $r(D_{\mathrm{eff}},T)$, evaluated at a reduced effective dimensionality $D_{\mathrm{eff}}<D$, rather than the field-free coefficient $r(D,T)$. This reduction, detailed below, underlies the phenomenon of ``dimensional depletion'' observed when a magnetic field is applied to a superconductor of nominal dimensionality $D$.

Assuming a strictly uniform flux density $\mathbf B=B\hat{\mathbf z}$, perpendicular to the CuO$_2$ planes ($x$--$y$), the orbital and spin degrees of freedom decouple. The corresponding Ginzburg--Landau Hamiltonian is then given by
\begin{equation}
\mathcal H_{\mathrm{GL}}[\varphi^*]=\mathcal H_n+\int d^Dr\Bigg[
\frac{r(D,T)}{2}|\varphi^*|^2+\frac b4|\varphi^*|^4
+\sum_{k=x,y}\frac{1}{2m_k^*}\left|\left(-i\hbar\partial_k-\frac{e^*}{c}A_k\right)\varphi^*\right|^2
+\frac{\hbar^2}{2m_z^*}|\partial_z\varphi^*|^2+\frac{B^2}{8\pi}\Bigg].
\label{eq:HGLfield}
\end{equation}
$\mathbf A$ denotes the vector potential, with $\mathbf B=\nabla\times\mathbf A$. The gradient of $\mathbf A$ is restricted to the $x$--$y$ plane; therefore, minimal coupling arises only in the in-plane terms. The term along $z$, the direction of $\mathbf B$ and coinciding with the crystallographic $c$-axis, remains a free kinetic term governed by the anisotropic mass $m_z^*$ introduced earlier. Since $|\varphi^*(\mathbf r)|$ is naturally expanded in the Landau basis $|N,k_z,q\rangle$, minimizing $\mathcal H_{\mathrm{GL}}$ yields the linearized eigenvalue equation
\begin{equation}
r(D,T)\varphi^*+\frac{\hbar^2}{2m^*}\left(-i\nabla_\perp-\frac{e^*}{\hbar c}\mathbf A\right)^2\varphi^*
+\frac{\hbar^2k_z^2}{2m_z^*}\varphi^*=E\varphi^*,
\label{eq:eigenvalue}
\end{equation}
solved with the standard ansatz $\varphi^*(\mathbf r)=\sum_qC_q\varphi_q^*$  \cite{Thouless, Gupta}. The linearized eigenvalue Eq.~\eqref{eq:eigenvalue} comprises three mutually commuting contributions to the spectrum, so they are directly additive. The term $r(D,T)$ represents a constant shift, independent of both position and momentum,
and functions as a uniform ``potential.'' The transverse minimal coupling term is equivalent to the standard 2D Landau problem; its eigenvalues are well established in both the symmetric and Landau gauges. The free term along the $z$-axis, diagonal in the plane-wave basis $e^{ik_zz}$, provides the standard kinetic energy. Because these three operators commute, the total eigenvalue is the sum of their individual contributions. Consequently, accounting for the Landau-level degeneracy per unit area, Eq.~\eqref{eq:eigenvalue} admits eigenvalues
\begin{equation}
E_{D,N,k_z}=r(D,T)+\left(N+\frac12\right)\hbar\omega_c+\frac{\hbar^2k_z^2}{2m_z^*}.
\label{eq:eigen}
\end{equation}
In this context, $N$ represents the Landau index, and $\omega_c=e^*B/m^*$ denotes the cyclotron frequency as defined in the introduction. Near the transition, only the lowest orbital ($N=0$) contributes significantly, as higher Landau levels are separated by $\hbar\omega_c$ and correspond to higher onset temperatures. By setting $N=0$ and $k_z=0$, this eigenvalue expression defines the field-dependent quadratic coefficient
\begin{equation}
r(D,T,B)=r(D,T)+\frac12\hbar\omega_c(B),
\label{eq:rDTB}
\end{equation}
and the mean-field critical condition $r(D,T,B)=0$ correspondingly determines the upper critical field,
\begin{equation}
r(D,T)=-\frac{\hbar e^*}{m^*}\,B_{c_2}(D,T).
\label{eq:Bc2}
\end{equation}
Because $r(D,T)$ incorporates the nonlinear, dimensionality-dependent structure derived previously, $B_{c_2}(D,T)$ and the corresponding field-dependent transition temperature $T_{c_B}$ exhibit non-monotonic behavior beyond the linear mean-field result. This non-monotonicity is well documented in Ginzburg--Landau analyses near quantum-critical points
and in superconductor-ferromagnet heterostructures, where competition between the ferromagnetic exchange field and superconducting pairing leads to a non-monotonic $T_{c_B}$ dependence on layer thickness.

\subsubsection{Fluctuation-specific heat in a magnetic field}

Thermal fluctuations of the OP generate an excess specific heat $C_{\text{fluc}}$ above the mean-field jump, with its precise form determined by the spatial dimensionality and the applied field. The dominant contribution at small $E_{D,N,k_z}$ is given by \cite{Thouless,Gupta}
\begin{equation}
C_{\text{fluc}} =\sum_{N,k_z}\frac{k_BT^2}{E_{D,N,k_z}^2}\left(\frac{dE_{D,N,k_z}}{dT}\right)^2.
\label{eq:Cfluc}
\end{equation}
In bulk (3D) samples with $\mathbf B$ applied along $z$, the transverse ($x$--$y$) degrees of freedom are fully quantized into Landau levels, leaving only free motion along $z$. The resulting fluctuations are effectively one-dimensional and are governed by $r(1,T)$ rather than $r(3,T)$, reflecting the dimensional reduction described above. Summing
over Landau levels and integrating over $k_z$,
\begin{equation}
C^{(3D)}_{\mathrm{fluc}}=\frac{k_BT^2}{2}\frac{e^*B}{2\pi\hbar}
\sum_{N=0}^\infty\int\frac{dk_z}{2\pi}
\left[\frac{r'(1,T)}{r(1,T)+\frac{\hbar^2k_z^2}{2m^*}+\hbar\omega_c(N+\frac12)}\right]^2,
\label{eq:C3D}
\end{equation}
where $r'(D,T)\equiv dr(D,T)/dT$ denotes the temperature derivative of the renormalized quadratic coefficient introduced earlier. In particular, $r'(1,T)$ depends on both temperature and the field-reduced effective dimensionality, serving as the dimension-dependent ``pseudo-temperature'' scaling variable also discussed above. Performing the $k_z$ integration explicitly,
\begin{equation}
C^{(3D)}_{\mathrm{fluc}}=k_BT^2\frac{e^*B}{4\pi\hbar}
\sum_{N=0}^\infty\frac{(2m^*)^{1/2}\,r'^2(1,T)}{\big[r(1,T)+\hbar\omega_c(N+1)\big]^{3/2}}.
\label{eq:C3Db}
\end{equation}
Reference~\cite{Thouless} extended these scaling relations above and below $T_{c_B}$, demonstrating that $C_{\mathrm{fluc}}$ transitions from Gaussian to critical scaling as the field modifies the effective density of states. Related dimensionality-specific behavior has been extensively documented  \cite{Quader,Gupta,Shenoy}. Physically, this 1D character reflects the extreme confinement of the pair wavefunction near the upper critical field. Orbital pair-breaking, mediated by the Lorentz force, restricts the transverse degrees of freedom available to Cooper pairs, while spin-paramagnetic pair-breaking (Zeeman splitting), not included explicitly in the present orbital-only Hamiltonian, provides an additional and complementary suppression mechanism.

In quasi-2D films with $\mathbf B$ applied perpendicular to the film, the $k_z$ degree of freedom is suppressed, confining carriers entirely to the lowest Landau level. The in-plane motion becomes fully localized into Landau orbits, resulting in 0D fluctuations governed by $r(0,T)$. The corresponding fluctuation-specific heat per unit area is
\begin{equation}
C^{(2D)}_{\mathrm{fluc}}=\frac{k_BT^2}{2}\frac{e^*B}{2\pi\hbar}
\sum_{N=0}^\infty\left[\frac{r'(0,T)}{r(0,T)+\hbar\omega_c(N+1)}\right]^2.
\label{eq:C2D}
\end{equation}
The transition temperature $T_{c_B}$ is defined by the vanishing of the minimum eigenvalue $E_{0,0}$ (i.e., $N=0$, $k_z=0$). Since $N=0$, Landau levels correspond to higher onset temperatures, the $N=0$ term dominates sufficiently close to $T_{c_B}$, where $\hbar\omega_c\gg r(D,T)$. In this limit, Eqs.~\eqref{eq:C3Db} and \eqref{eq:C2D} reduce to
\begin{align}
C^{(3D)}_{\mathrm{fluc}}&=k_BT^2\frac{e^*B}{4\pi\hbar}
\frac{(2m^*)^{1/2}\,r'^2(1,T)}{\big[r(1,T)+\hbar\omega_c\big]^{3/2}},
\label{eq:C3Dlimit}\\
C^{(2D)}_{\mathrm{fluc}}&=\frac{k_BT^2}{2}\frac{e^*B}{2\pi\hbar}
\left[\frac{r'(0,T)}{r(0,T)+\hbar\omega_c}\right]^2.
\label{eq:C2Dlimit}
\end{align}

Collectively, these results demonstrate that applying a magnetic field to a nominally $D$-dimensional superconductor reduces the effective dimensionality of the fluctuation problem to $D-2$ (bulk 3D becomes effectively 1D; thin-film 2D becomes effectively 0D) by quantizing the degrees of freedom transverse to $\mathbf B$ into discrete Landau levels. This field-induced dimensional reduction, expressed through $r(D_{\mathrm{eff}},T)$ and its temperature derivative, provides a unified framework for understanding how orbital pair-breaking modifies both the upper critical field $B_{c_2}(D,T)$ and the fluctuation-driven specific heat in layered cuprates such as YBCO.

\section{Numerical results and discussion}
\label{sec:numerics}

\begin{figure}[htbp]
\centering
\includegraphics[width=0.32\linewidth]{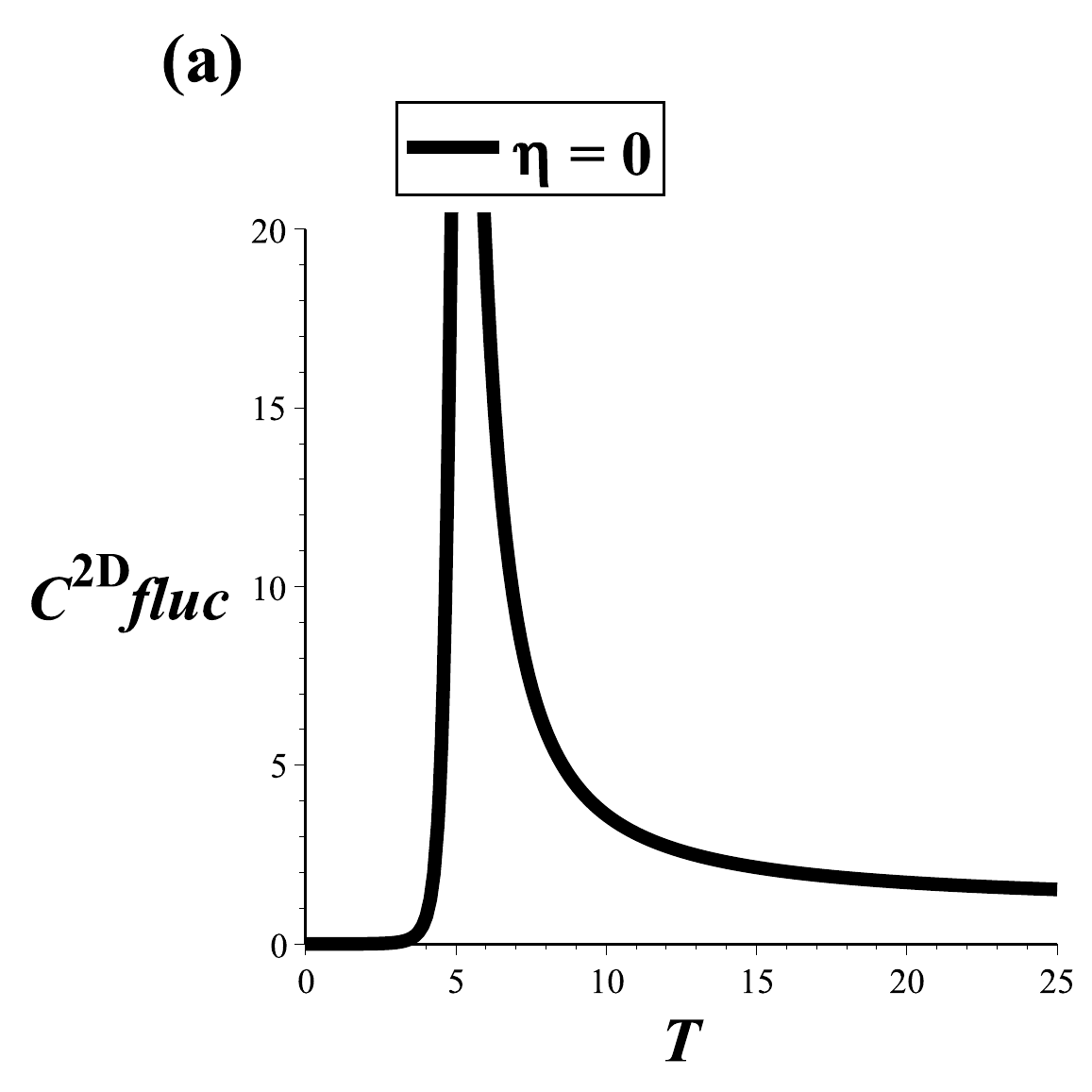}
\includegraphics[width=0.32\linewidth]{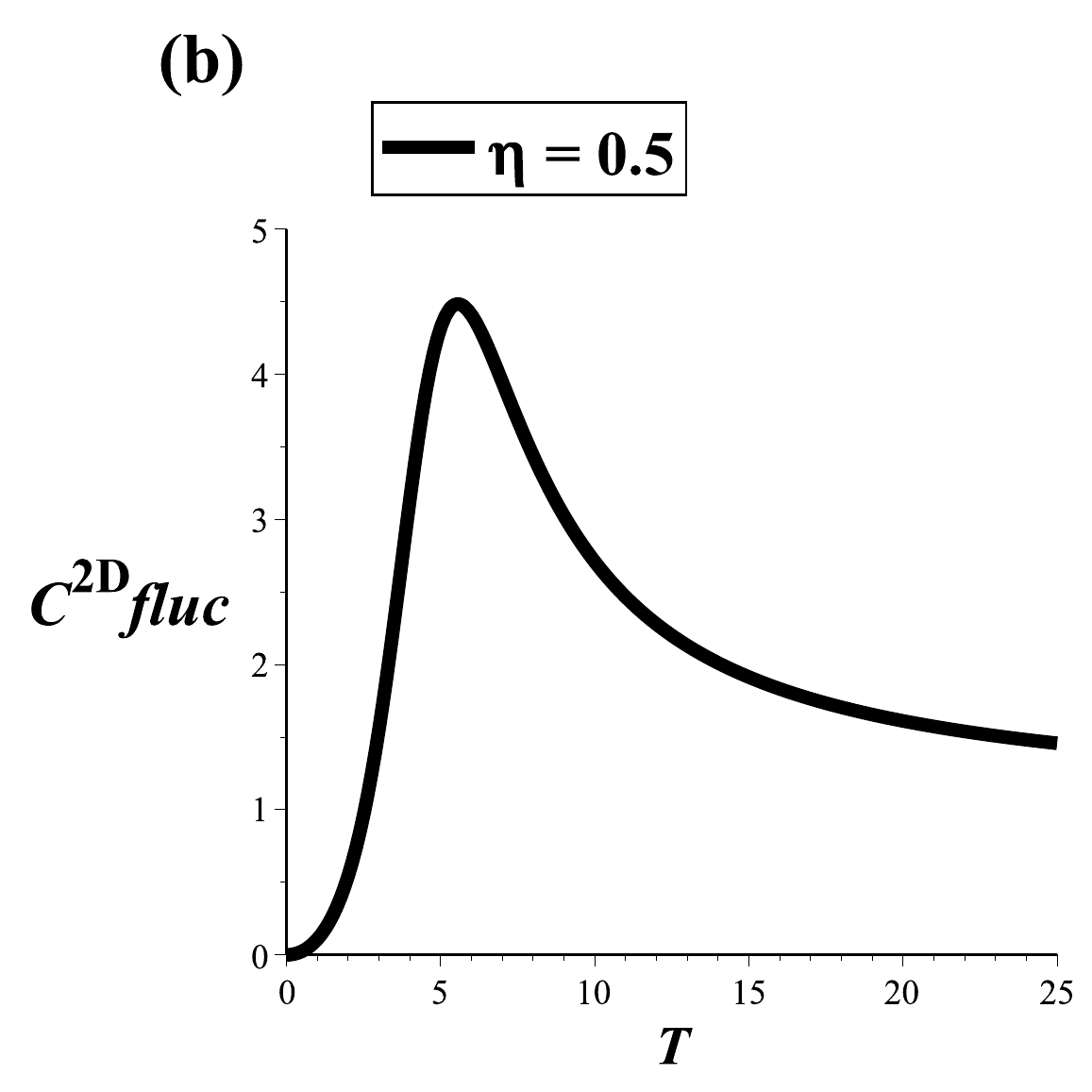}
\includegraphics[width=0.32\linewidth]{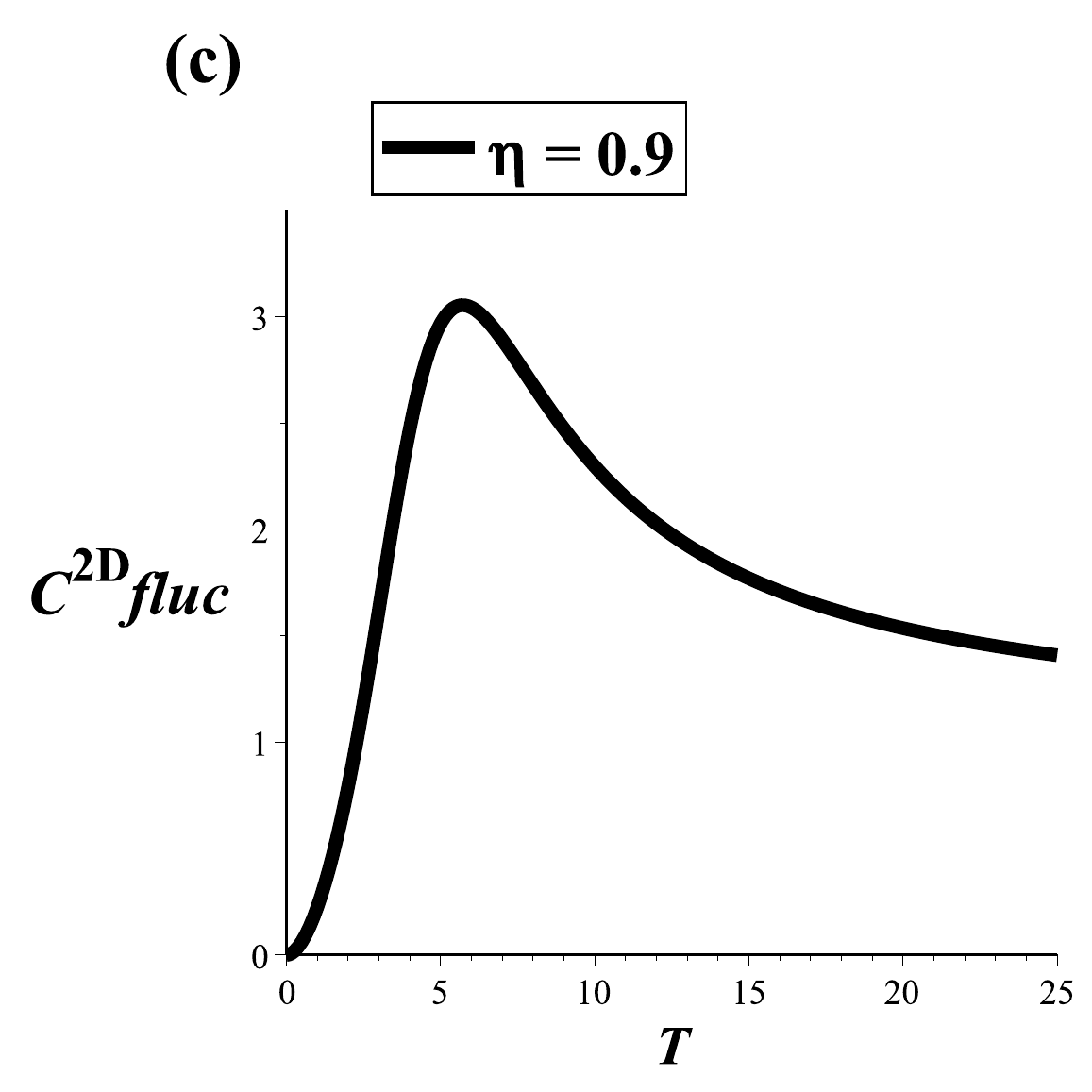}
\caption{2D fluctuation-specific heat at a fixed magnetic field ($B=2.85$~T) as the fluctuation parameter $\eta$ is varied: (a) $\eta=0$; (b) $\eta=0.5$; (c) $\eta=0.9$ ($\hbar\omega_c=0.33$~meV throughout). Application of a magnetic field suppresses superconductivity, thereby exposing the normal state and an extended regime of superconducting
fluctuations. In contrast to the mean-field divergence anticipated in clean 2D systems for $\eta=0$, the fluctuation-specific heat $C^{2D}_{\mathrm{fluc}}$ typically broadens without diverging and approaches the normal-state background at elevated temperatures.}
\label{fig:1}
\end{figure}

\begin{figure}[htbp]
\centering
\includegraphics[width=0.32\linewidth]{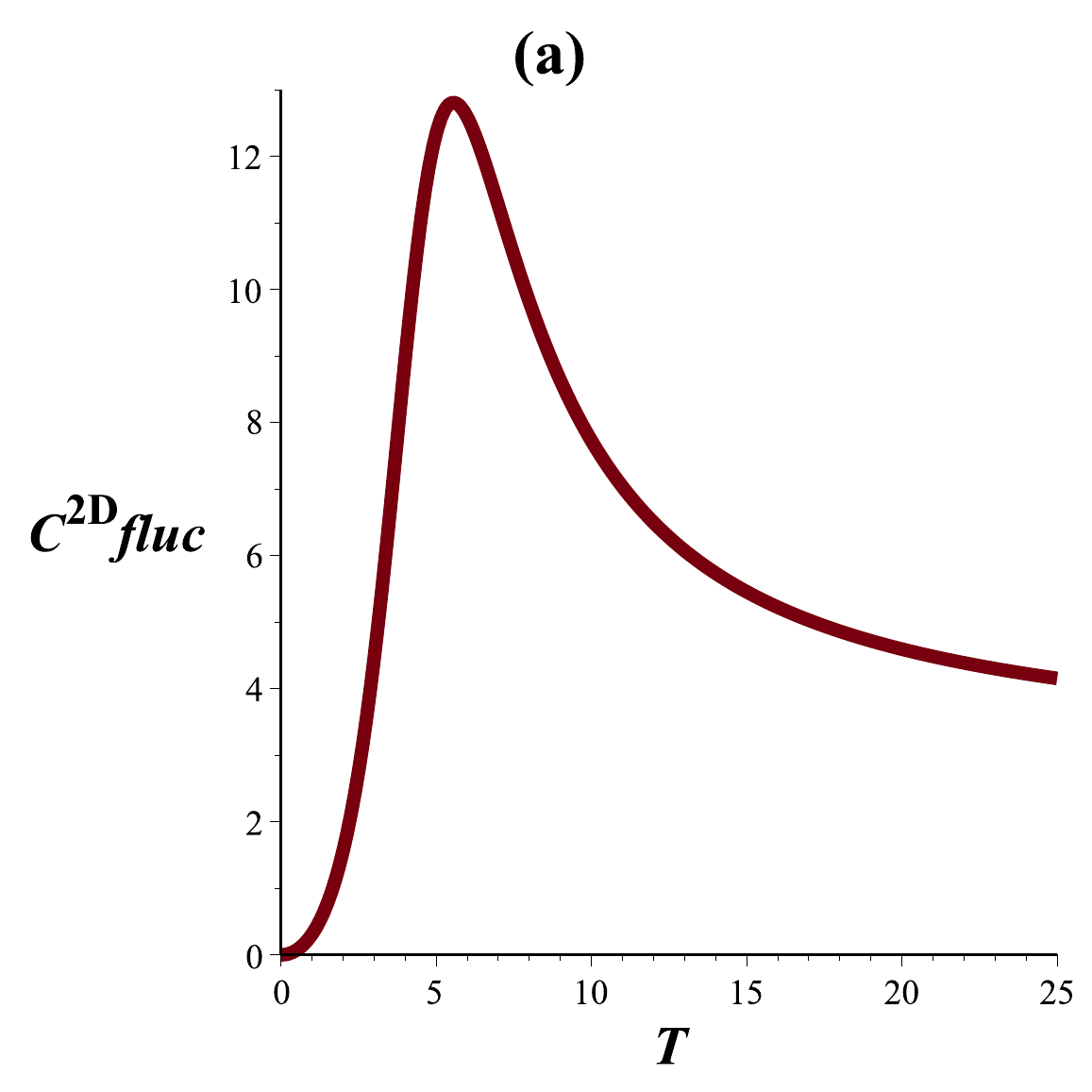}
\includegraphics[width=0.32\linewidth]{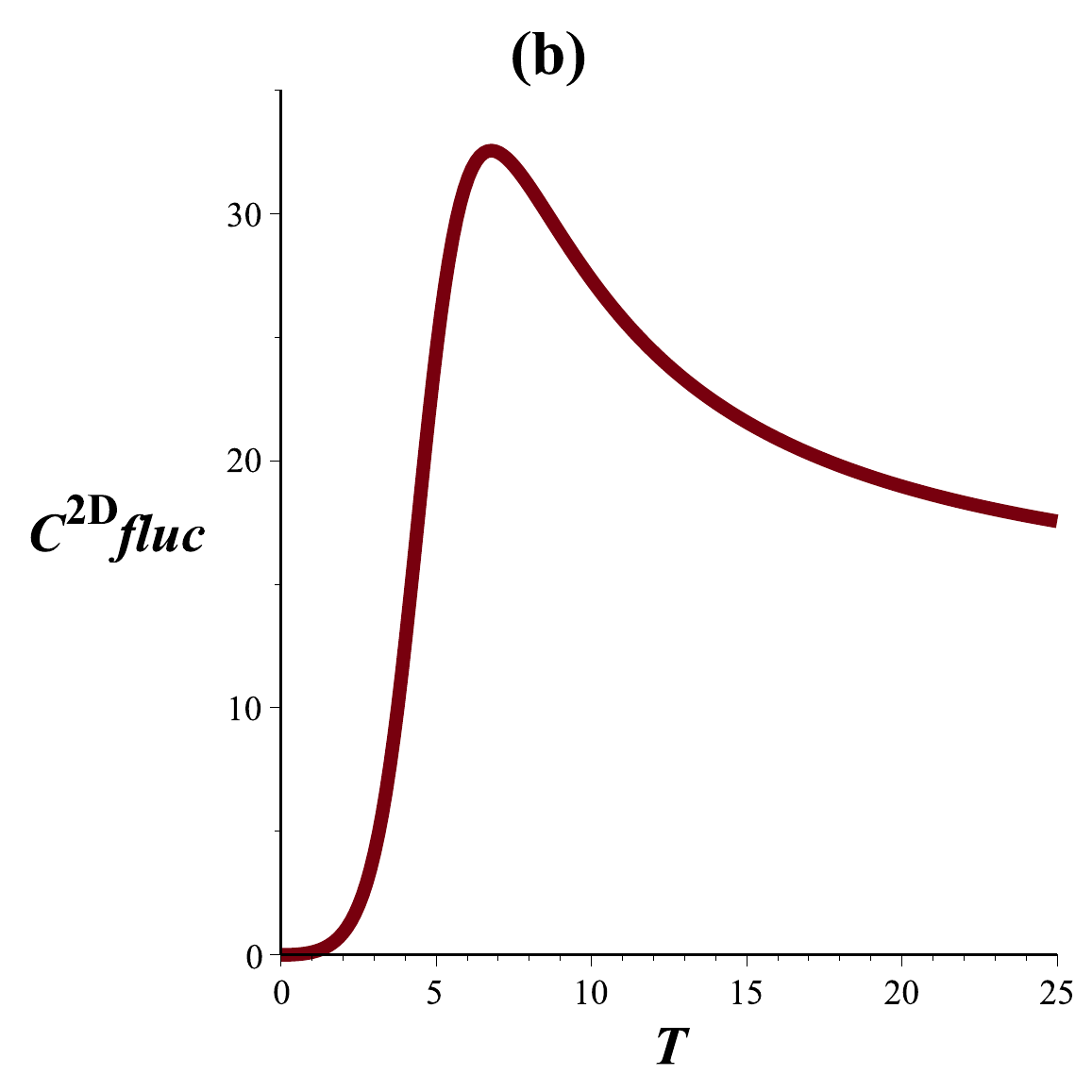}
\includegraphics[width=0.32\linewidth]{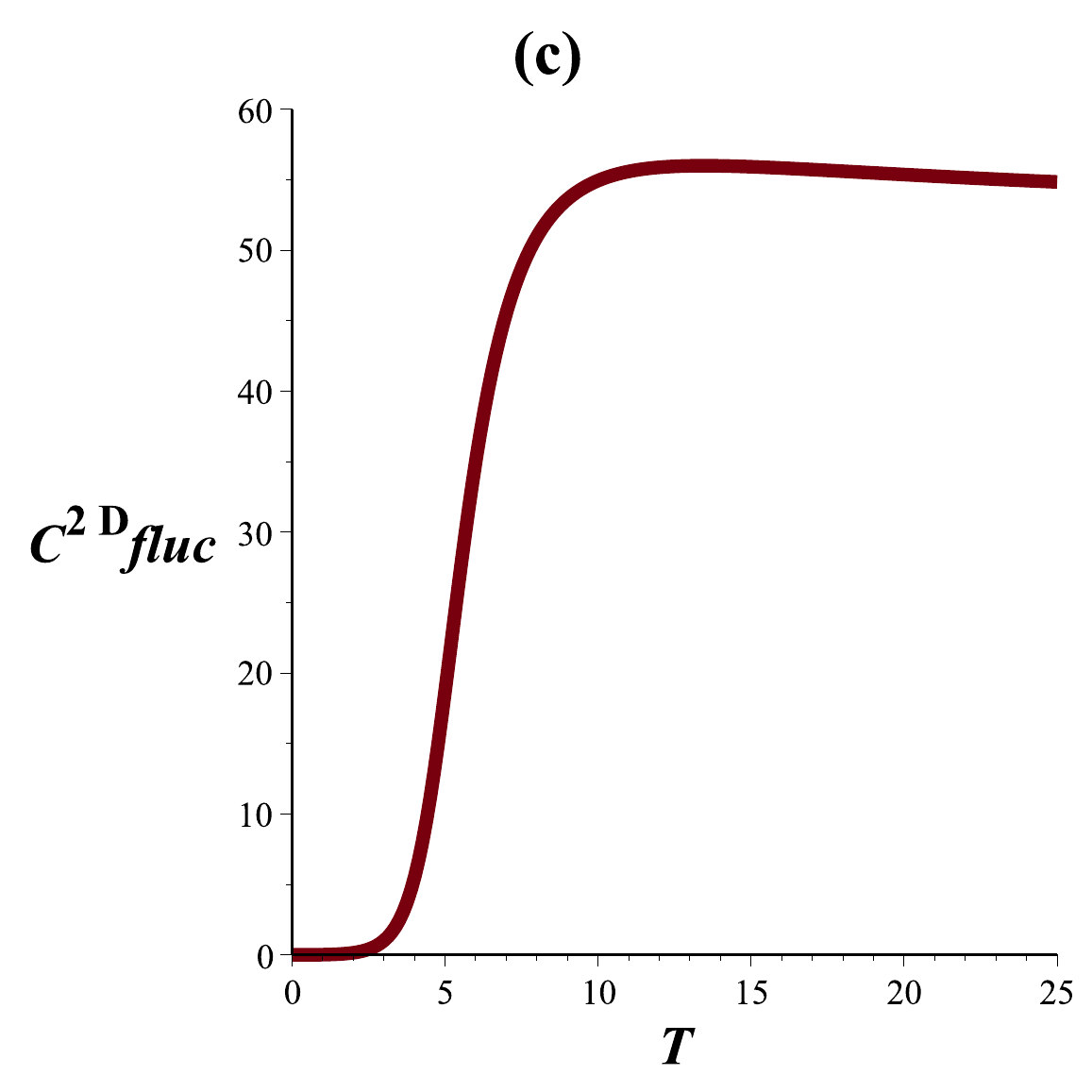}
\caption{2D fluctuation-specific heat at a fixed fluctuation parameter $\eta=0.5$ as the magnetic field is varied: (a) $\hbar\omega_c=0.33$~meV, $B=2.85$~T; (b) $\hbar\omega_c=1.5$~meV, $B=12.96$~T; (c) $\hbar\omega_c=3$~meV, $B=25.92$~T. An external magnetic field breaks Cooper pairs, weakens superconductivity, and reveals thermodynamic fluctuation effects in 2D systems.}
\label{fig:2}
\end{figure}

\begin{figure}[htbp]
\centering
\includegraphics[width=0.32\linewidth]{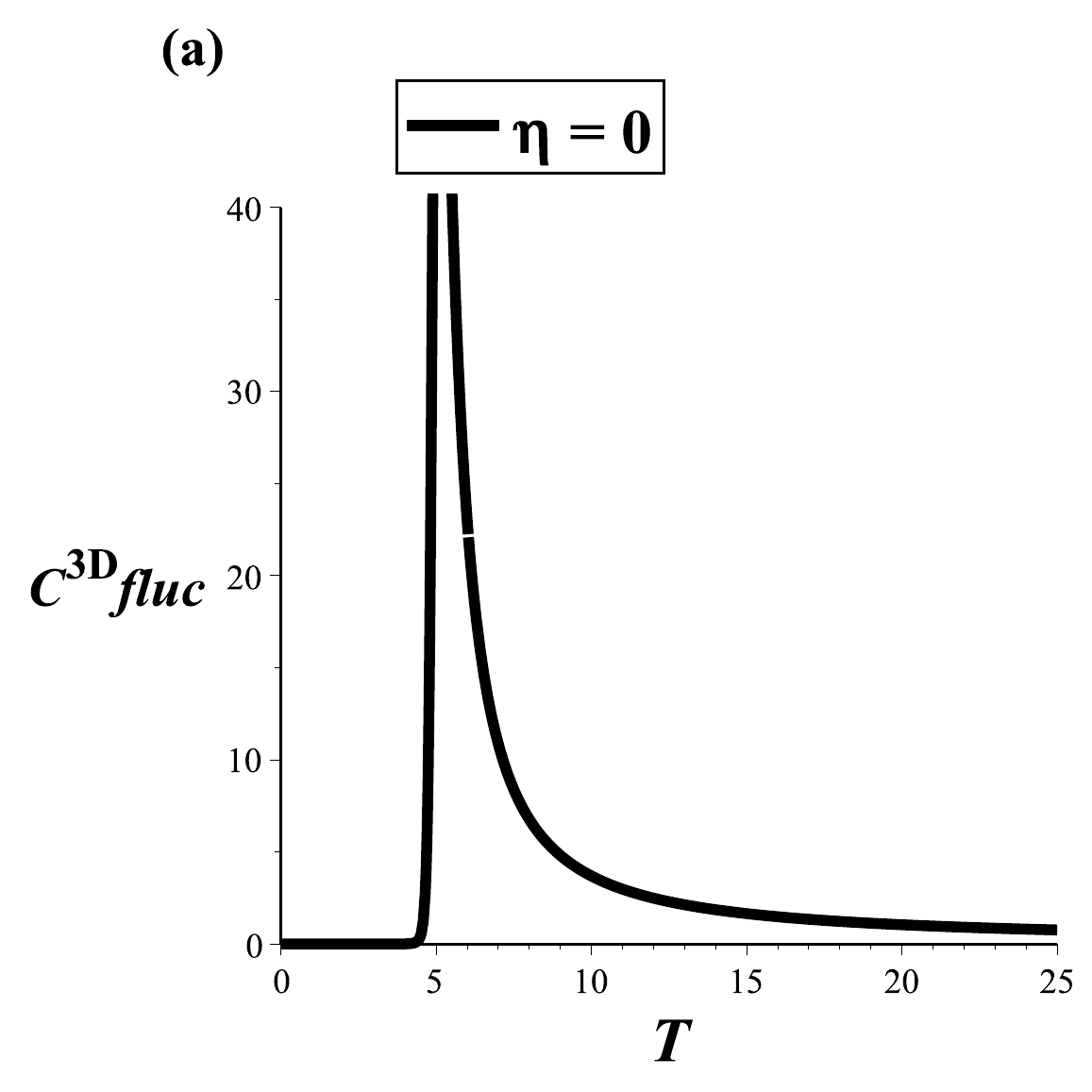}
\includegraphics[width=0.32\linewidth]{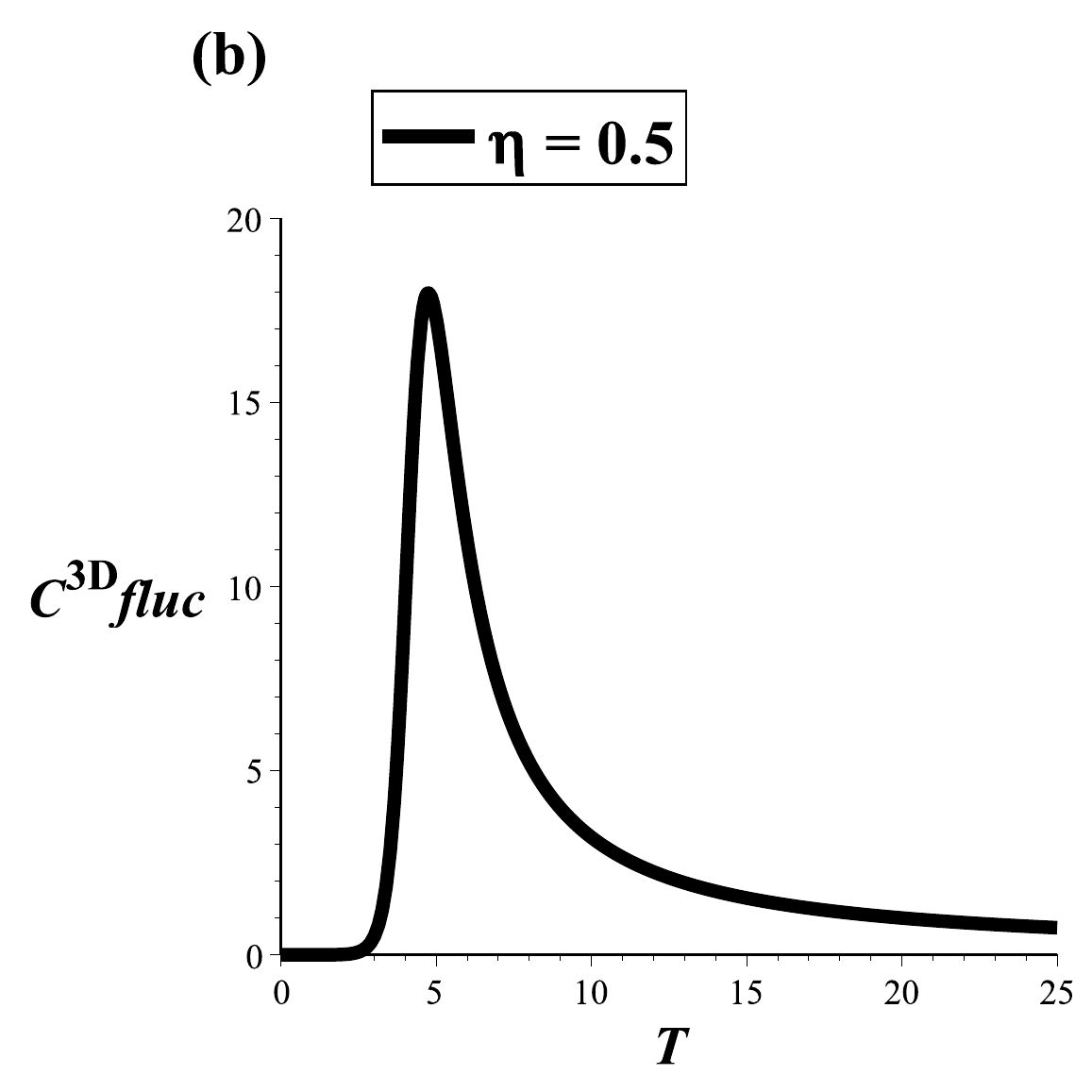}
\includegraphics[width=0.32\linewidth]{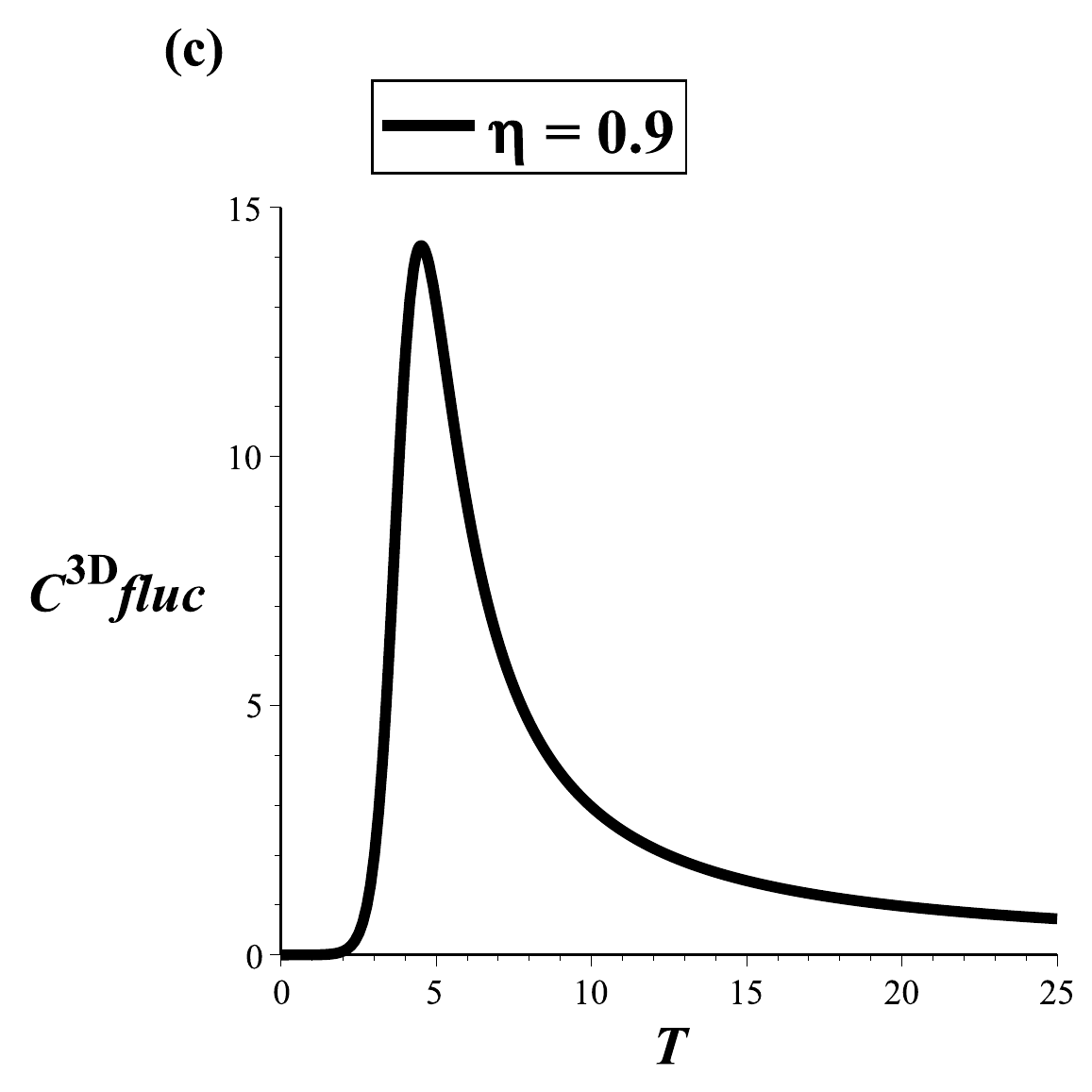}
\caption{3D fluctuation-specific heat at a fixed magnetic field ($B=12.96$~T) as the fluctuation parameter $\eta$ varies: (a) $\eta=0$; (b) $\eta=0.5$; (c) $\eta=0.9$ ($\hbar\omega_c=1.5$~meV throughout). Unlike the mean-field divergence expected for $\eta=0$ in clean 3D systems, the fluctuation-specific heat $C^{3D}_{\mathrm{fluc}}$ usually
broadens and approaches the normal-state background at higher temperatures.}
\label{fig:3}
\end{figure}

\begin{figure}[htbp]
\centering
\includegraphics[width=0.32\linewidth]{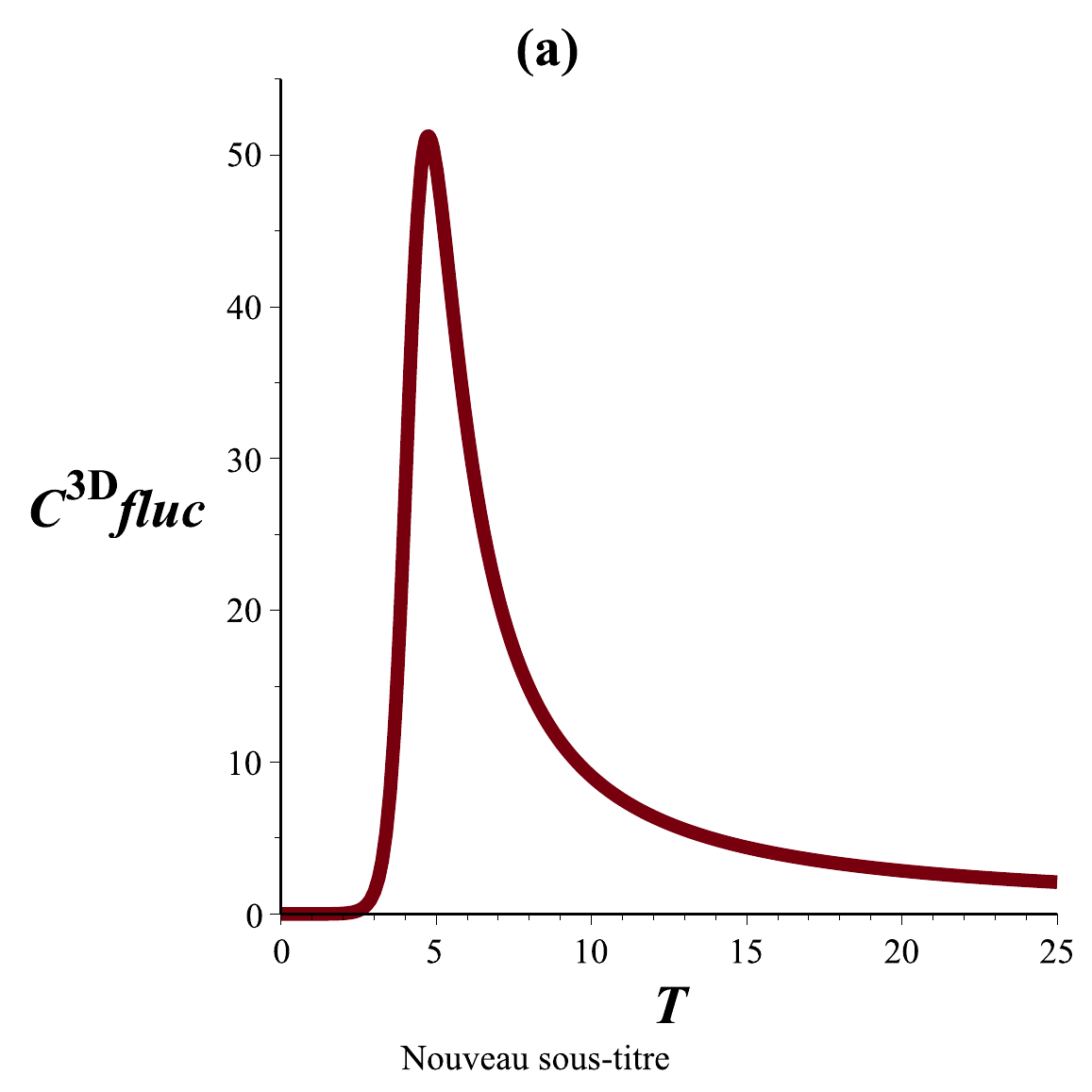}
\includegraphics[width=0.32\linewidth]{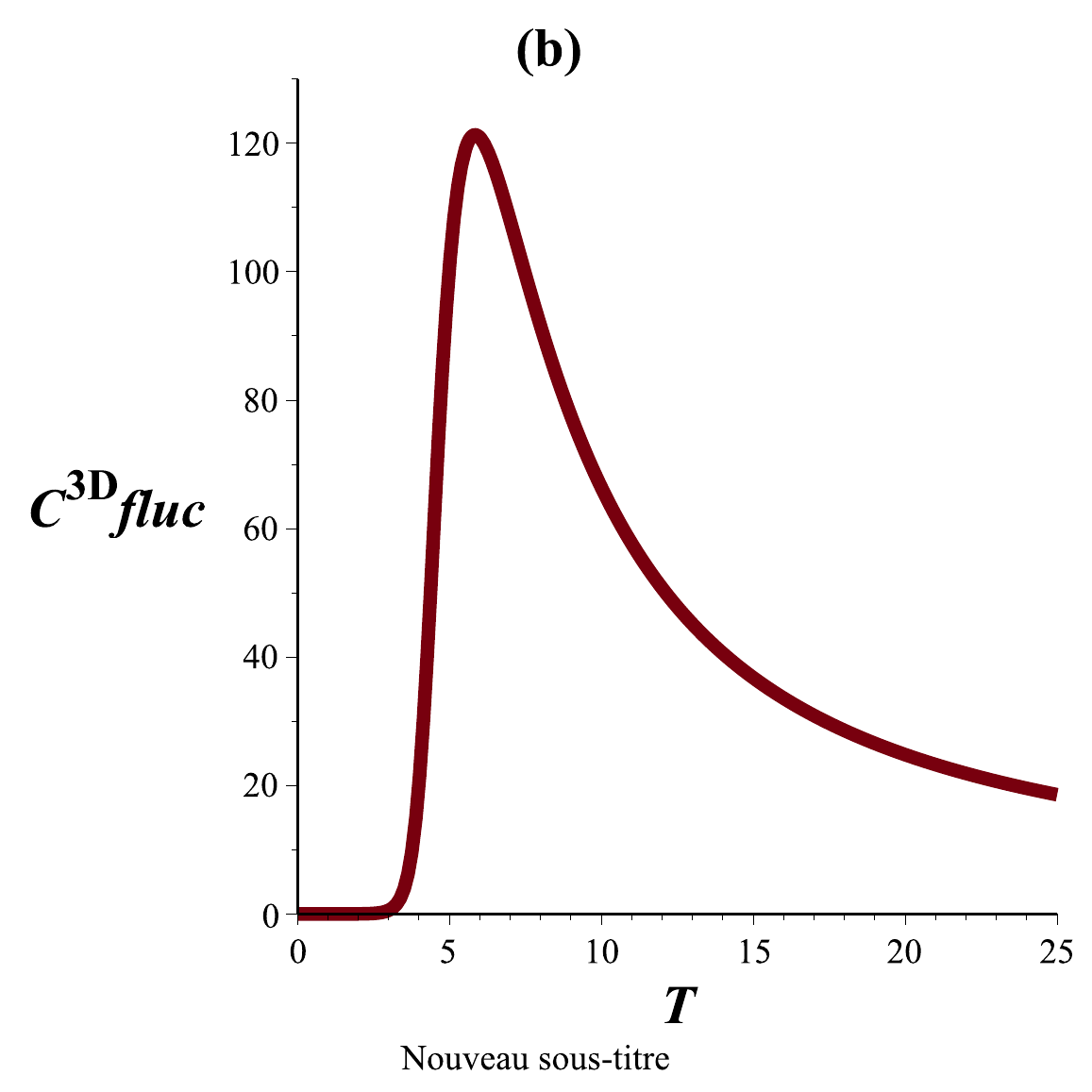}
\includegraphics[width=0.32\linewidth]{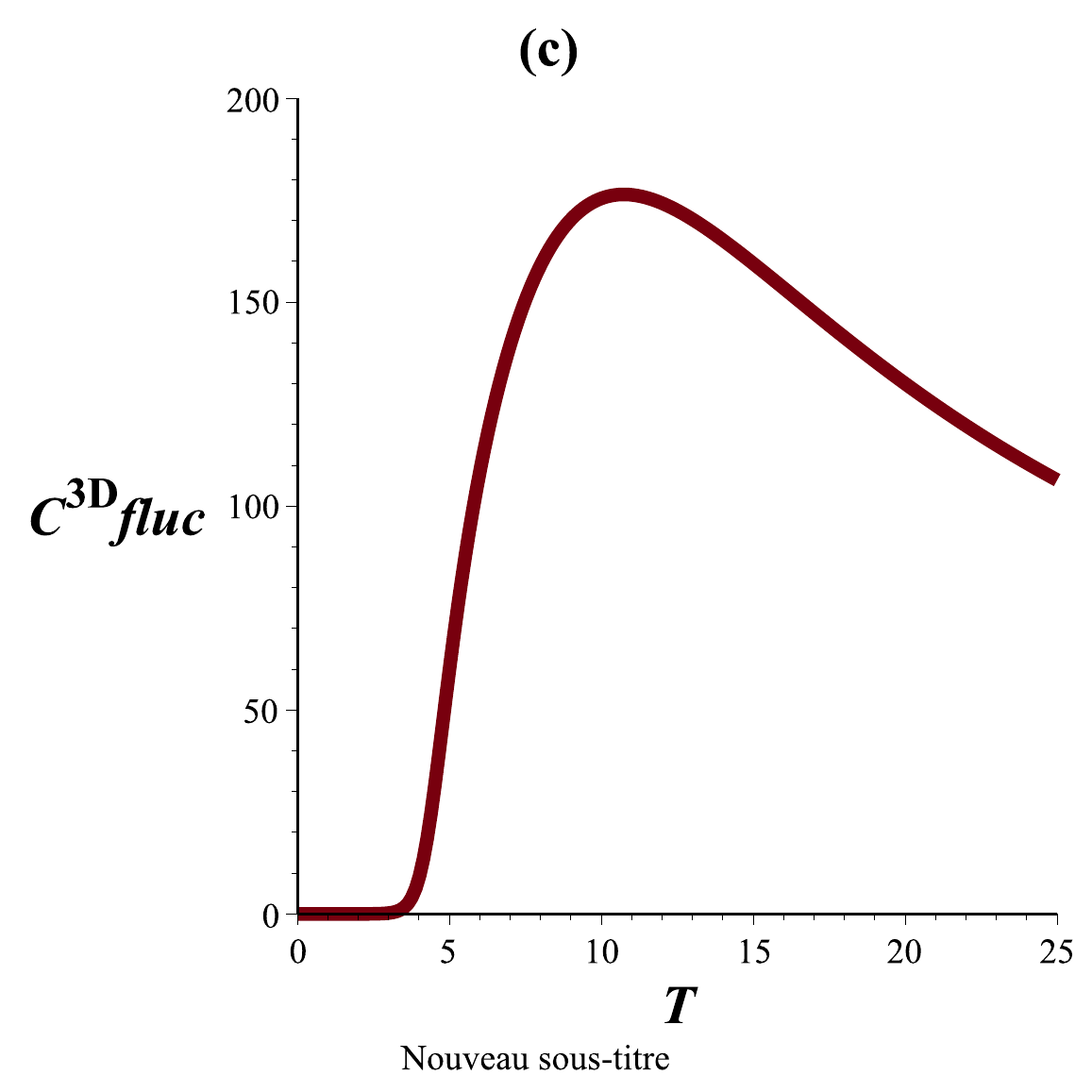}
\caption{3D fluctuation-specific heat at a fixed fluctuation parameter $\eta=0.5$ as the magnetic field varies: (a) $\hbar\omega_c=0.33$~meV, $B=2.85$~T; (b) $\hbar\omega_c=3$~meV, $B=25.92$~T; (c) $\hbar\omega_c=10$~meV, $B=86.38$~T. A magnetic field breaks Cooper pairs, weakens superconductivity, and reveals 3D thermodynamic
fluctuations, exposing the normal state and broadening the fluctuation regime.}
\label{fig:4}
\end{figure}

This section plots the fluctuation specific heat as a function of the fluctuation coupling $\eta$, the applied magnetic field $B$, and temperature $T$. A low-dimensional superconductor with a transition temperature ($T_{c_0}$) near 5~K often exhibits pronounced phase fluctuations, dimensional crossovers, or collective quantum phenomena that distinguish it from conventional 3D bulk materials. (This idealized $T_{c_0}=5$~K scale is used in Figs.~\ref{fig:1}--\ref{fig:4} to illustrate the general shape of the theory before the model is anchored to the physical YBCO transition temperature $T_{c_0} \approx 92$--$93$~K in Sec.~\ref{sec:experiment}.)

As established above, $\eta$ and $B$ enter the theory on the same footing, through the combined field- and fluctuation-renormalized coefficient  $r(D,T,B)=r(D,T)+\tfrac12\hbar\omega_c$: together, they determine the thermodynamic stability of the superconducting state, which is lost precisely when $r(D,T,B)$ changes sign. When either thermal/quantum fluctuations (large $\eta$) or the applied field (large $B$, via $\omega_c$) grow sufficiently large relative to the bare condensation scale $r_0T_{c_0}$, this sign change, rather than a discontinuous mean-field transition, is smoothed into the extended, fluctuation-dominated crossover discussed throughout this work.

Figures~\ref{fig:1}--\ref{fig:4} present the fluctuation specific heat as a function of temperature for different values of the material-specific coupling $\eta$, allowing direct comparison between the present model and the conventional mean-field treatment. The interplay between low dimensionality and the upper critical field is further examined by comparing the lowest-Landau-level energy $\tfrac12\hbar\omega_c$ against the renormalized coefficient $r(D_{\mathrm{eff}},T)$, in three limiting regimes. In the first (low-field) regime, $\tfrac12\hbar\omega_c\ll r(D_{\mathrm{eff}},T)$, corresponding to weak applied fields; the field enters only as a small perturbation, and the fluctuation specific heat closely follows its field-free, $D_{\mathrm{eff}}$-dimensional form. In the second, crossover regime, $\tfrac12\hbar\omega_c\sim r(D_{\mathrm{eff}},T)$:
the two energy scales are comparable, and neither the field-free nor the fully field-dominated limit provides an adequate description, requiring the full expressions derived above. In the third, high-field regime, $\tfrac12\hbar\omega_c\gg r(D_{\mathrm{eff}},T)$, characteristic of fields approaching $B_{c2}(D_{\mathrm{eff}},T)$, the $N=0$ Landau level dominates and the reduced expressions [Eqs.~\eqref{eq:C3Dlimit}--\eqref{eq:C2Dlimit}] apply, reflecting the field-induced reduction to an effectively lower
dimensionality.

The present approach renormalizes the amplitude of the fluctuation-specific heat through the coupling $\eta$, extending the conventional treatment by explicitly incorporating order-parameter fluctuations and particle scattering, while preserving the underlying critical exponents set by the (effective) dimensionality $D_{\mathrm{eff}}$. In this sense, the renormalized scaling curves obtained here differ quantitatively, in overall magnitude and field/temperature dependence, from those of conventional treatments that neglect this dimensionality dependence, without altering the universality class itself. In particular, the resulting curve shapes in 2D- and 3D systems under an applied field are markedly distinct, reflecting the characteristic, discretized behavior expected of a system with finite (Landau-quantized) energy levels.

Figure~\ref{fig:1} presents data for a fixed $B=2.85$~T and varying values of $\eta=0,0.5,0.9$. The $\eta=0$ curve serves as a robust validation, reproducing the sharp, near-divergent mean-field-like spike at $T_{c_0} =5$~K, consistent with the prediction of Eq.~\eqref{eq:rDT} when fluctuation coupling is absent. Introducing nonzero values of $\eta$ eliminates the divergence and reduces the peak amplitude ($\eta=0.5$ to $\eta=0.9$). This figure provides the strongest support for the central claim that $\eta$, independent of the magnetic field, regularizes the divergence.

Figure~\ref{fig:2} presents data for a fixed $\eta=0.5$, with $\hbar\omega_c$ varied from 0.33 to 3~meV. The amplitude increases substantially with increasing field (approximately 13 to 35 to 60). At the highest field, curve (c) no longer resembles a peak; instead, it saturates into a plateau rather than decaying. This represents a qualitative change in the curve shape, not merely a scaling effect, and aligns with the ``high-field regime'' described in the text, where the $N=0$ Landau level dominates, and $r(0,T)+\hbar\omega_c$ remains essentially field-locked over the plotted range.

For $\eta=0$, both $r(1,T)$ and $r(0,T)$ vanish at $T_{c_0} =5$~K. Figures~\ref{fig:1}(a) and \ref{fig:3}(a) display a singularity or discontinuity in the fluctuation-specific heat at this temperature, which indicates a continuous phase transition characterized by divergent microscopic energy fluctuations and correlation lengths, rather than a
finite mean-field jump. This result demonstrates convergence between the present approach and mean-field theory. For $\eta>0$, both $r(1,T)$ and $r(0,T)$ vanish at $T_c=0$~K. The results exhibit distinct profiles with properties characteristic of finite-energy systems, as illustrated in Figs.~\ref{fig:1}(b), \ref{fig:1}(c), \ref{fig:3}(b), and \ref{fig:3}(c). Incorporation of the fluctuation factor $\eta$ smooths abrupt discontinuities and divergences. Further increases in $\eta$ round sharp edges and reduce the amplitude of the jump. The combined effects of fluctuations and very low magnetic fields prohibit the formation of an ordered phase at $T_c=0$~K, as shown in Figs.~\ref{fig:2}(a) and \ref{fig:4}(a). Temperature-independent behavior at low temperatures is dominated by quantum-fluctuation dominance. The fluctuation-specific heats
($C^{2D}_{\mathrm{fluc}}$ and $C^{3D}_{\mathrm{fluc}}$) display pronounced anharmonic behavior at low temperatures driven by quantum fluctuations, but cross over to constant values when a low magnetic field is applied. This crossover is attributed to the opening of a Zeeman or excitation gap, which suppresses low-energy thermal and quantum modes. This behavior stands in contrast to predictions from standard GLT.

Figure~\ref{fig:3} presents data for a fixed magnetic field of $B=12.96$~T, with $\eta$ varying from 0 to 0.9. This figure extends the analysis of Fig.~\ref{fig:1} into three dimensions. A divergence is observed at $\eta=0$. Direct comparison of Figs.~\ref{fig:1} and \ref{fig:3} indicates that the attenuation effect associated with $\eta\ne0$ in three dimensions is proportionally weaker than in two dimensions. This result is consistent with the physical interpretation that one-dimensionally confined fluctuations, which govern the 3D case under a field via $r(1,T)$, are less restrictive than the zero-dimensionally confined fluctuations characteristic of the 2D case.

Figure~\ref{fig:4} presents data for a fixed value of $\eta=0.5$ with $\hbar\omega_c$ varying from 0.33 to 10~meV. The amplitude increases sharply with the field (approximately 50 to 130 to 180), and the peak shifts to higher temperatures with noticeable broadening. This trend is qualitatively similar to that observed in Fig.~\ref{fig:2}, although the
absolute field range is larger.

An increase in the applied field, which leads to higher Landau levels, significantly alters the nature of fluctuations. This change subsequently influences the width of the critical region due to enhanced quantum degeneracy and reduced fluctuation effects at higher Landau levels. Specifically, in the 2D case the critical region narrows with increasing Landau level, yet remains substantially below the renormalized quadratic coefficient within the temperature range near $T_{c_0}$, where a decrease in specific heat is observed [Fig.~\ref{fig:2}(c)]. In contrast, in the 3D case this region widens with increasing Landau level, remaining well above the renormalized quadratic coefficient [Fig.~\ref{fig:4}(c)]. It is necessary to ensure that this widening does not become so pronounced as to undermine the validity of the previous analysis in the temperature range where a significant reduction in specific heat is anticipated. The behavior of the critical region at higher Landau levels thus depends on the dimensionality: it narrows as fluctuations weaken in 2D and widens, while remaining bounded to maintain analytical validity near $T_{c_0}$, in 3D. The interaction between low-dimensional systems and magnetic fields produces highly sensitive and tunable dynamic behavior, which contrasts significantly with the behavior observed in bulk systems.

The renormalized approach incorporates strong thermal fluctuations and layered anisotropy into the standard mean-field theory. For cuprates in a magnetic field, this framework characterizes the transition from Abrikosov vortex lattices to a vortex liquid and evaluates the impact of critical fluctuations on the upper critical field boundary. The analysis is formulated using Eq.~\eqref{eq:Bc2}, where the upper critical field is directly linked to the renormalized quadratic coefficients. Within this framework, the upper critical field becomes a function of the system's dimensionality, resulting in non-monotonic dynamic behavior. Replacing $r(D,T)+\hbar\omega_c$ with $[T-T_{c_B}]r'(D,T)$ enables the fluctuation-specific heat to be expressed as a function of the upper critical field. Including the temperature derivative of the renormalized quadratic coefficient introduces a new dynamic, leading to a dependence on $(T-T_{c_0})^\varepsilon$, where the exponent $\varepsilon$ is specific to the system dimensionality. Fluctuations alter the mean-field specific-heat exponents and induce a crossover between distinct fixed points as the correlation length $\xi$ increases relative to the layer thickness or
confinement length. The absence of divergence in the fluctuation-specific heat indicates that the superconducting transition is either rounded out or smeared into a crossover. Instead of a sharp, diverging fluctuation-specific heat, thermal fluctuations and vortices destroy the long-range phase coherence, decoupling pairing amplitude from phase
stiffness. In layered cuprates, intense magnetic fields suppress the 3D Ginzburg--Landau coherence, rendering the fluctuations effectively 1D. This dimensional restriction prevents the thermodynamic divergence typically expected in bulk materials. Instead of diverging at the critical temperature $T_c$, the fluctuation-specific heat exhibits a rounded, continuous enhancement.

\begin{table}[htbp]
\centering
\caption{Parameter selection for constructing the YBa$_2$Cu$_3$O$_{7-\delta}$ dimensional crossover.}
\label{tab:1}
\begin{tabular}{lc}
\hline
Quantity & Value adopted \\
\hline
$T_{c_0}$ & 92~K \\
$\xi_{ab}(0)$ (in-plane coherence length) & 1.35~nm \\
$\rho=\xi_{ab}/\xi_c$ (mass anisotropy) & 6 \\
$\xi_c(0)=\xi_{ab}(0)/\rho$ & 0.225~nm \\
$\Delta C_0$ & 45~mJ~mol$^{-1}$~K$^{-2}$ \\
$d\approx c/2$ (interplanar spacing) & 0.6~nm \\
\hline
\end{tabular}
\end{table}

\section{Comparison of experimental and theoretical results for YBa$_2$Cu$_3$O$_{7-\delta}$}
\label{sec:experiment}

In YBa$_2$Cu$_3$O$_{7-\delta}$ (YBCO) with $0\le\delta\le0.18$, variations in oxygen stoichiometry influence the density of states at the Fermi level and, consequently, the magnitude of the specific-heat jump $\Delta C_p$ at the superconducting transition  \cite{Keumo3, Cooper, DFisher}. Two intrinsic material parameters render YBCO a particularly suitable testbed for this framework. First, its bare zero-temperature coherence length, $\xi_0\simeq1$~nm, is approximately three orders of magnitude smaller than the $\xi_0\simeq1000$~nm typical of conventional BCS superconductors. This places YBCO within the regime where the Ginzburg criterion predicts an experimentally accessible fluctuation-dominated region, in contrast to the extremely narrow window observed in conventional superconductors. Second, its quasi-2D CuO$_2$-layered structure implies that the effective dimensionality relevant to $r(D,T)$ is not strictly determined by the nominal 3D crystal geometry, but instead transitions between two and three dimensions depending on the relationship between the coherence length along the $c$-axis and the interlayer spacing. Both factors contribute to making dimensionality-dependent fluctuation corrections sufficiently large to be observable, rather than constituting a minor perturbation to an otherwise sharp mean-field transition.

\begin{figure}[htbp]
\centering
\includegraphics[width=0.55\linewidth]{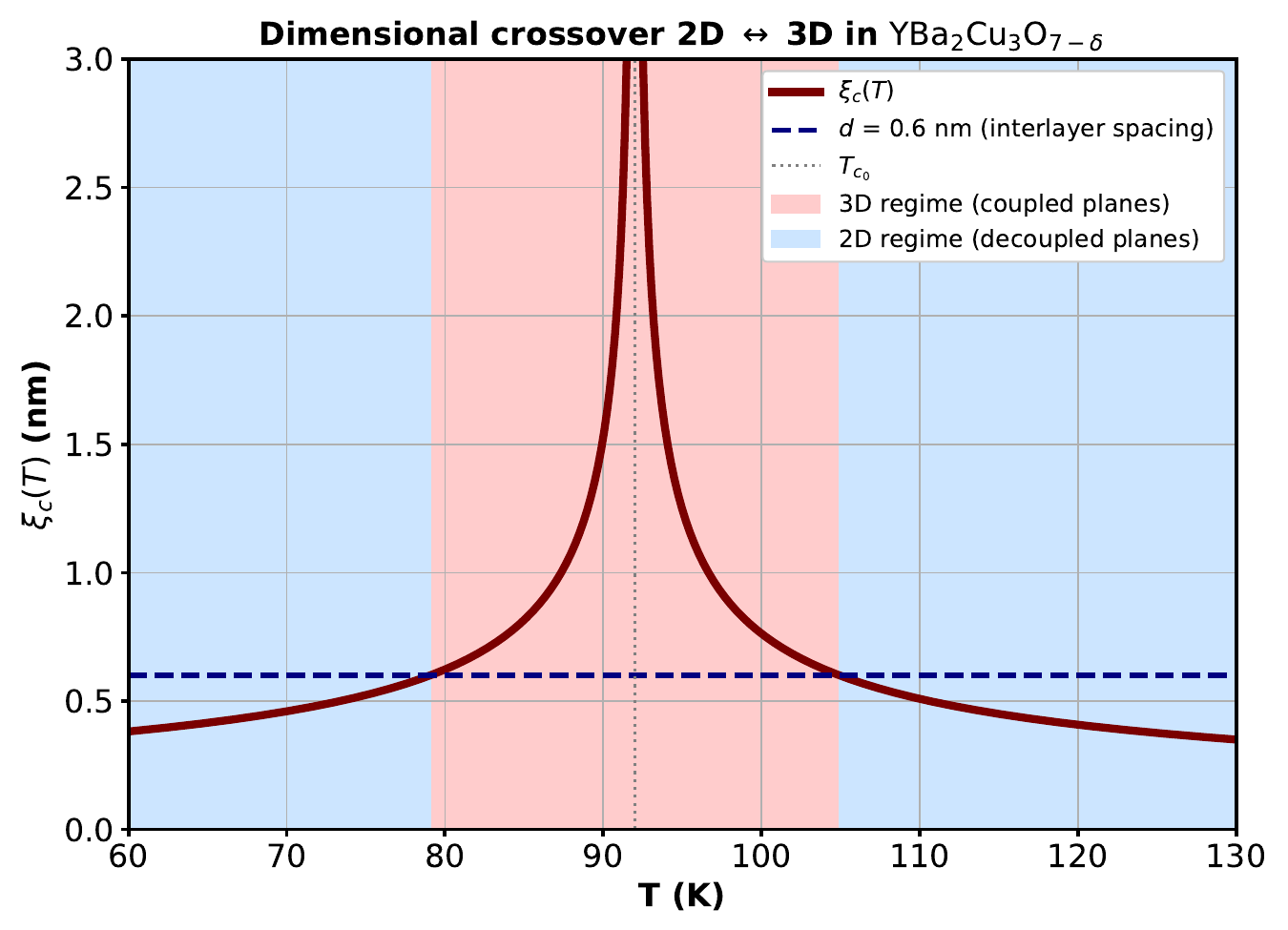}
\caption{Development of 2D and 3D configurations for the YBa$_2$Cu$_3$O$_{7-\delta}$ dimensional crossover: perpendicular coherence length $\xi_c(T)$ versus temperature, compared with the interlayer spacing $d=0.6$~nm.}
\label{fig:5}
\end{figure}

The transition from 2D to 3D behavior in YBCO is determined by the relationship between the perpendicular coherence length,
\begin{equation}
\xi_c(T)=\xi_c(0)|\tau|^{-1/2},\qquad \tau=\frac{T-T_{c_0}}{T_{c_0}},
\label{eq:xic}
\end{equation}
and the interplanar spacing $d$. For $\xi_c(T)>d$ (near $T_{c_0}$), the CuO$_2$ planes are coupled, resulting in a 3D regime. For $\xi_c(T)<d$ (far from $T_{c_0}$), the planes decouple, leading to a 2D regime as described by the Aslamazov--Larkin model. Consistent selection of the values listed in Table~\ref{tab:1} enables the construction of 2D and 3D configurations for the dimensional crossover in YBCO.

Figure~\ref{fig:5} illustrates a central result of the formalism: a single length-scale comparison, specifically the perpendicular coherence length $\xi_c(T)$ versus the interlayer spacing $d$, suffices to explain why YBa$_2$Cu$_3$O$_{7-\delta}$ exhibits both 3D-XY critical behavior and 2D Aslamazov--Larkin fluctuation behavior, depending solely on the proximity of $T$ to $T_{c_0}$. The curve shows that $\xi_c(T)=\xi_c(0)|\tau|^{-1/2}$ diverges symmetrically as $T$ approaches $T_{c_0}$ from either side, consistent with mean-field coherence-length expectations. The intersection of the curve with the horizontal line $d=0.6$~nm at two symmetric points, $T^\star_{\mathrm{low}}\approx79$~K and
$T^\star_{\mathrm{high}}\approx105$~K, represents the substantive physical content of the plot, rather than the divergence itself (which is a mean-field artifact regularized in practice by the fluctuation corrections developed in $r(D,T)$ \cite{Keumo3}). The 3D (red) band is intentionally narrower than the 2D (blue) band; this asymmetry conveys the principal qualitative insight. It indicates that a YBCO sample is ``genuinely 3D'' only within a narrow temperature range around $T_{c_0}$, while a much broader adjacent region exhibits predominantly 2D physics. This interpretation aligns with experimental specific-heat data on YBCO, such as the sharp, narrow 3D-XY cusp superimposed on a broader rounded background, as discussed in the Phillips--Fisher--Gordon review  \cite{Phillips}. The figure thus provides a concrete geometric explanation for this experimental observation.

In the absence of a magnetic field, these two physical configurations correspond directly to the two previously established branches: in three dimensions, $r(3,T)$, or its branch consistent with mean-field theory, is relevant exclusively within the narrow (3D-band) temperature range. In two dimensions, the quadratic coefficient is defined as
$r(2,T)=r_0(T-T_{c_0})+\eta T/T_{c_0}$. This formulation is valid across the 2D-band region and beyond, where $\varepsilon=\eta/(r_0T_{c_0})>0.2$, indicating the predominance of 2D Aslamazov--Larkin fluctuations.

The parameter $\eta$ remains constant between the two configurations; instead, the effective dimensionality $D$ changes, governed by the comparison between $\xi_c(T)$ and $d$. This mechanism of dimensional reduction, previously discussed for the magnetic-field-induced crossover, is applied here to an intrinsic, temperature-driven crossover specific to
the lamellar structure of YBCO. The calculated crossover does not represent a system that becomes 2D far from $T_{c_0}$. Rather, it describes a fundamentally 3D system with extreme anisotropy, whose behavior mimics 2D characteristics over an intermediate temperature range. This phenomenon arises because a genuine 2D system cannot produce the observed transition, as the Mermin-Wagner theorem establishes.

\begin{figure}[htbp]
\centering
\includegraphics[width=0.32\linewidth]{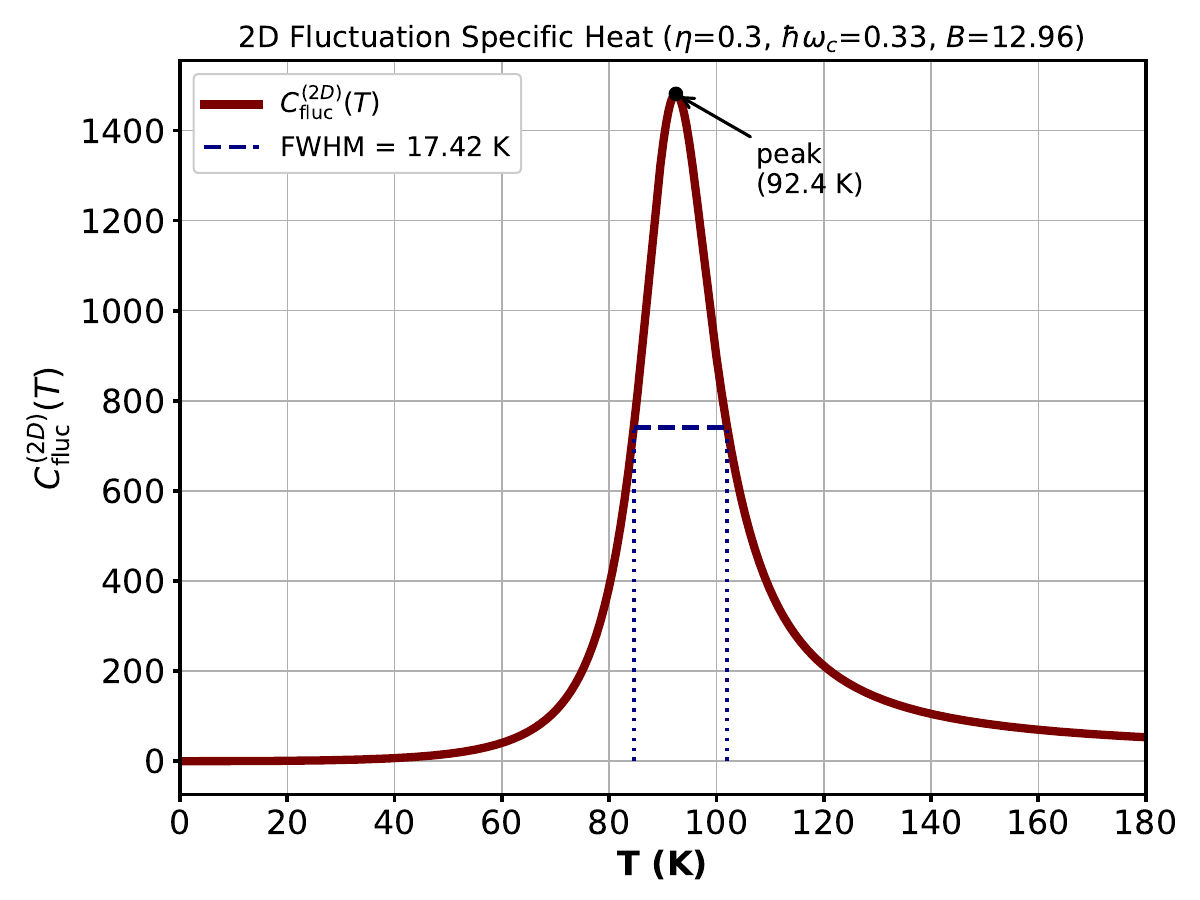}
\includegraphics[width=0.32\linewidth]{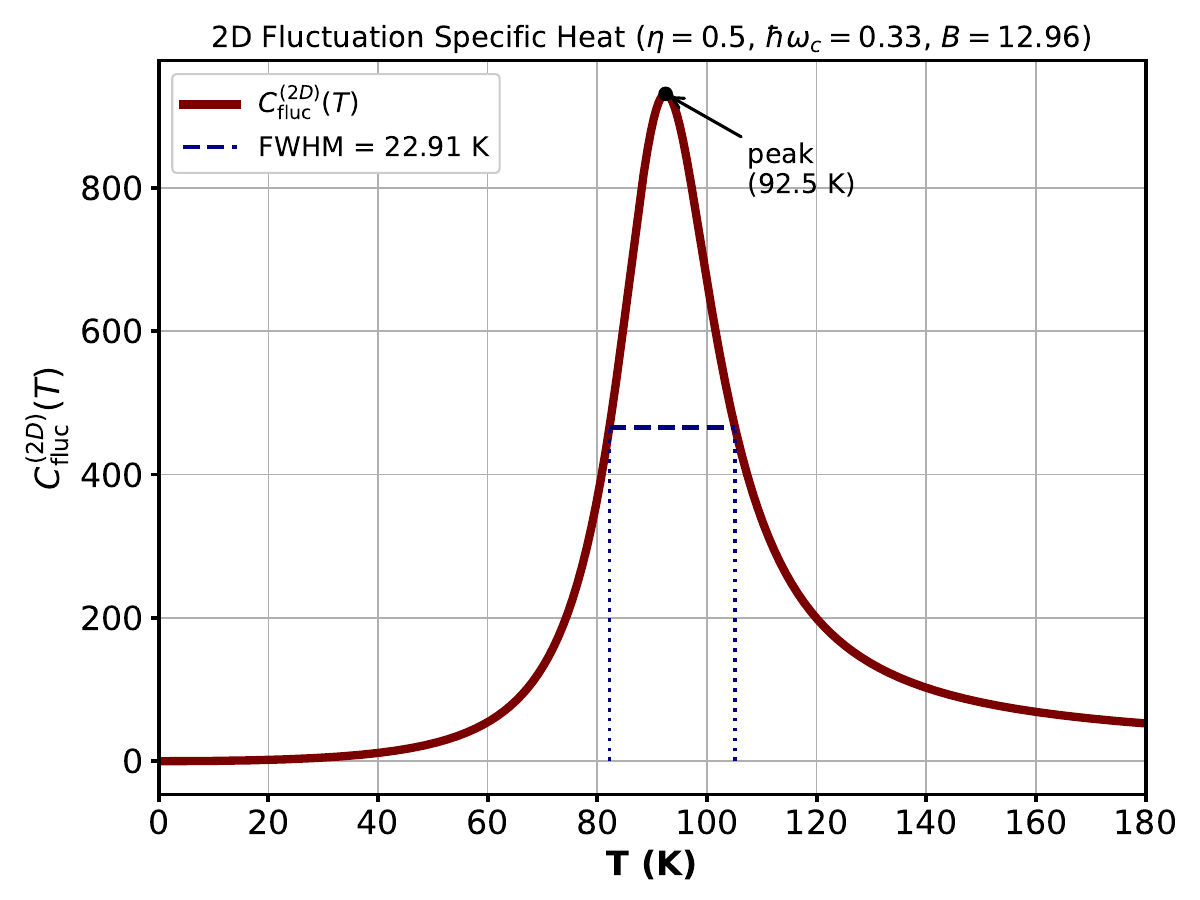}
\includegraphics[width=0.32\linewidth]{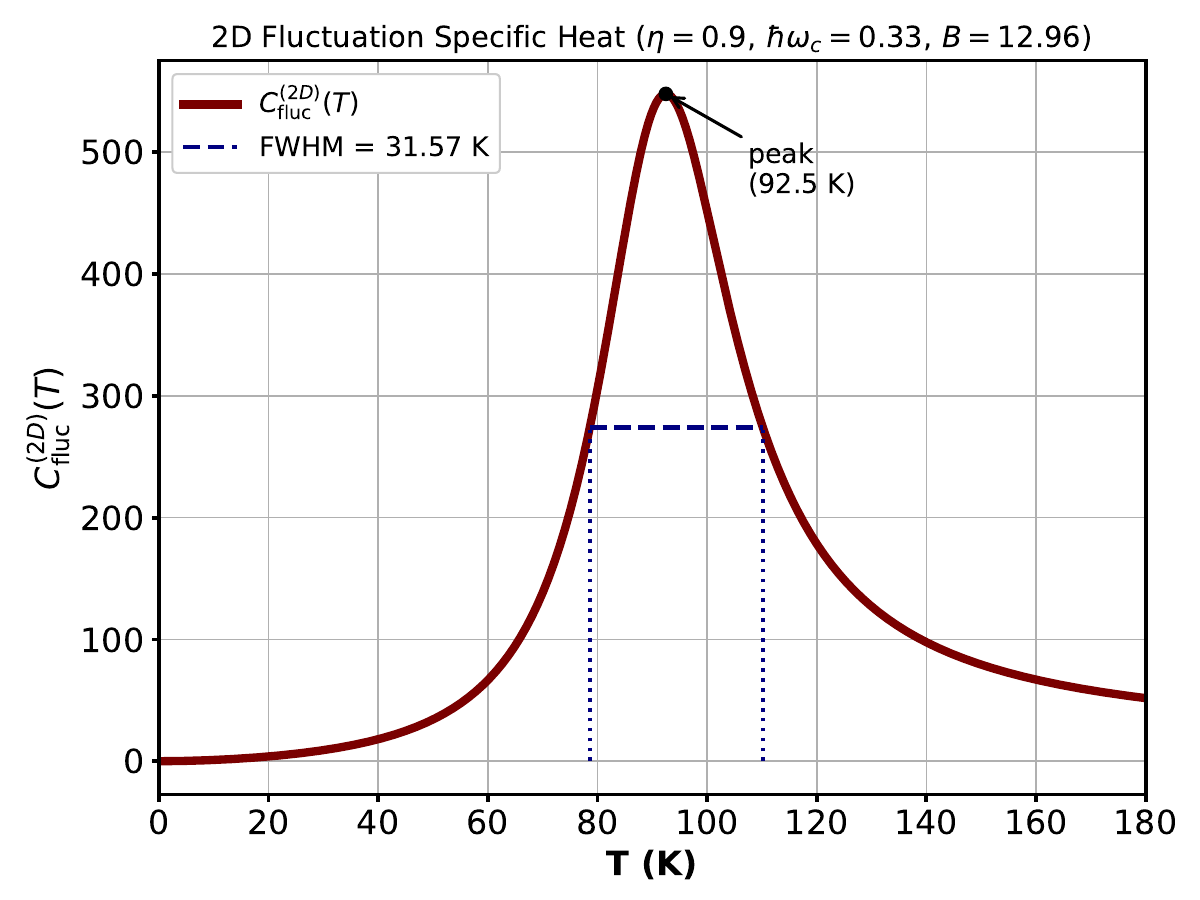}
\caption{2D fluctuation-specific heat at a fixed value of the magnetic field $B=2.85$~T ($\hbar\omega_c=0.33$~meV) as the fluctuation parameter $\eta$ is varied: (a) $\eta=0.3$, FWHM$=17.42$~K; (b) $\eta=0.5$, FWHM$=22.91$~K; (c) $\eta=0.9$, FWHM$=31.57$~K. The amplitude of the fluctuation-specific heat decreases with increasing $\eta$.}
\label{fig:6}
\end{figure}

\begin{figure}[htbp]
\centering
\includegraphics[width=0.32\linewidth]{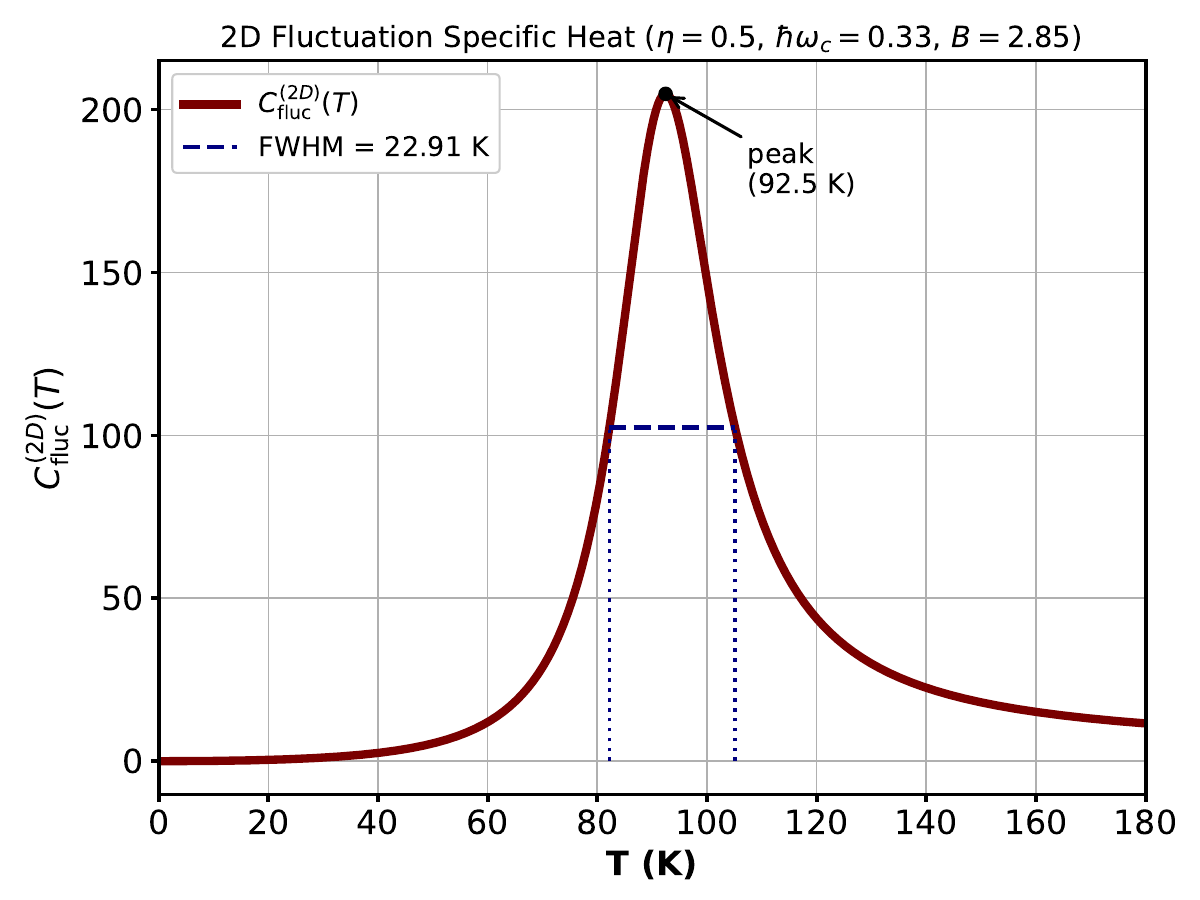}
\includegraphics[width=0.32\linewidth]{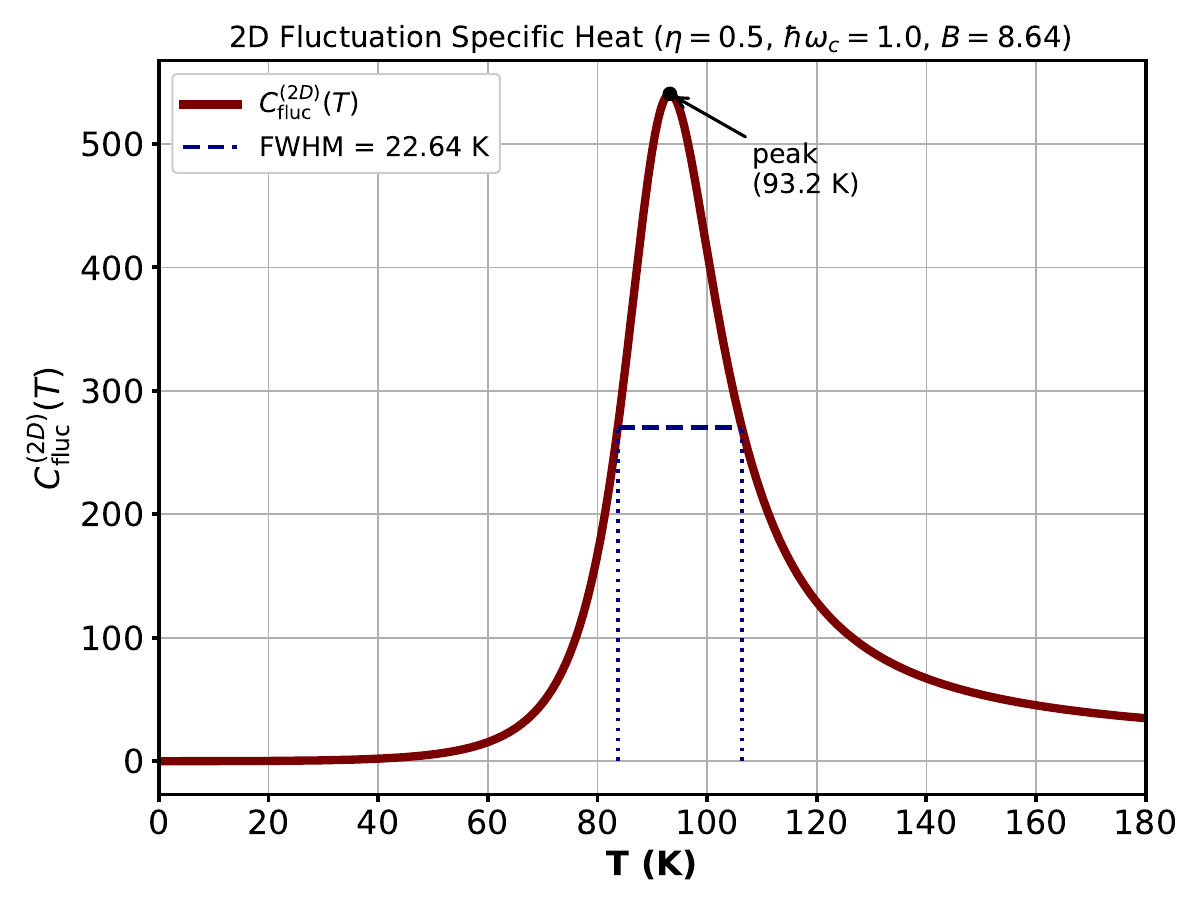}
\includegraphics[width=0.32\linewidth]{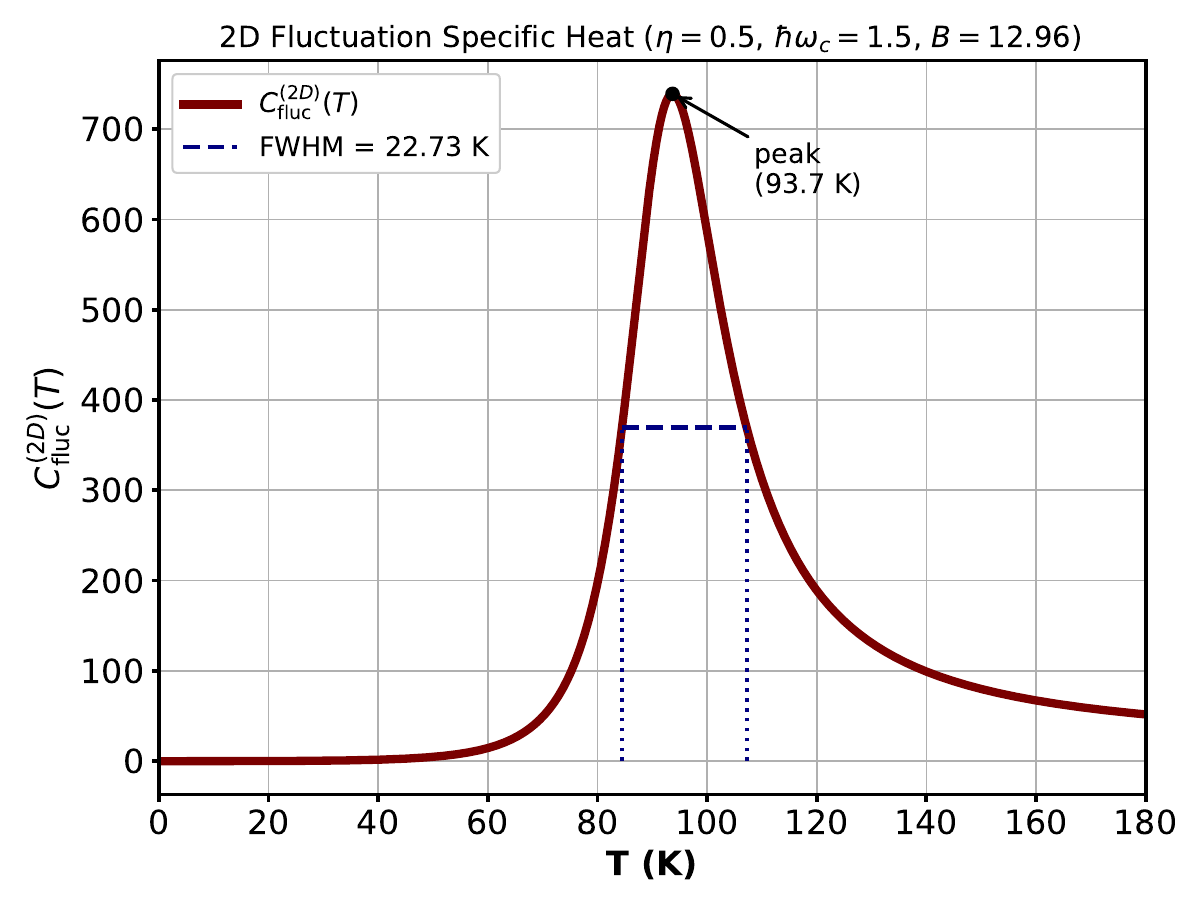}
\caption{2D fluctuation-specific heat at a fixed value of the fluctuation
parameter $\eta=0.5$ as the magnetic field is varied: (a) $\hbar\omega_c=0.33$~meV, $B=2.85$~T, FWHM$=22.91$~K; (b) $\hbar\omega_c=1$~meV, $B=8.64$~T, FWHM$=22.64$~K; (c)
$\hbar\omega_c=1.5$~meV, $B=12.92$~T, FWHM$=22.73$~K. The amplitude of the fluctuation-specific heat increases with increasing magnetic field; higher fields suppress both thermal and spatial order-parameter fluctuations, which stabilizes the system and reduces excess thermodynamic contributions.}
\label{fig:7}
\end{figure}

\begin{figure}[htbp]
\centering
\includegraphics[width=0.32\linewidth]{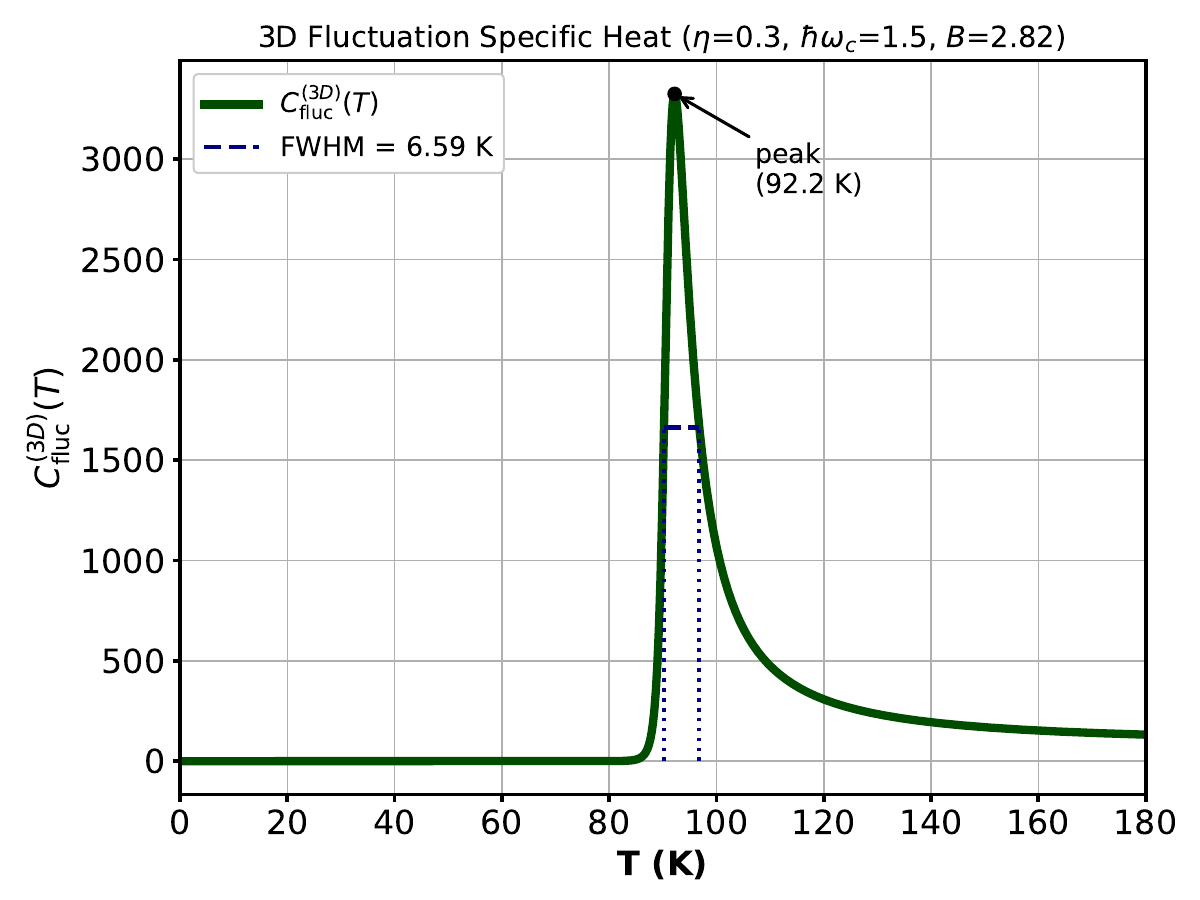}
\includegraphics[width=0.32\linewidth]{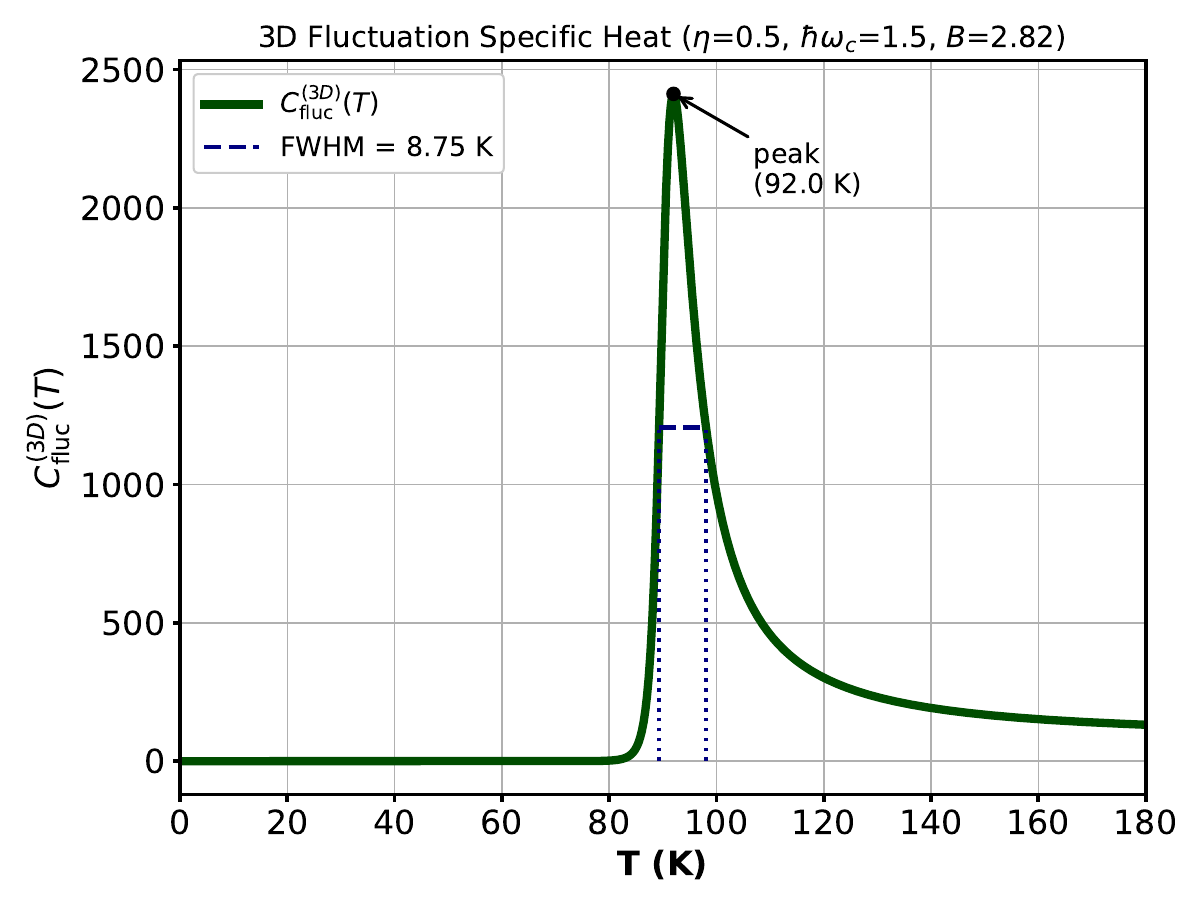}
\includegraphics[width=0.32\linewidth]{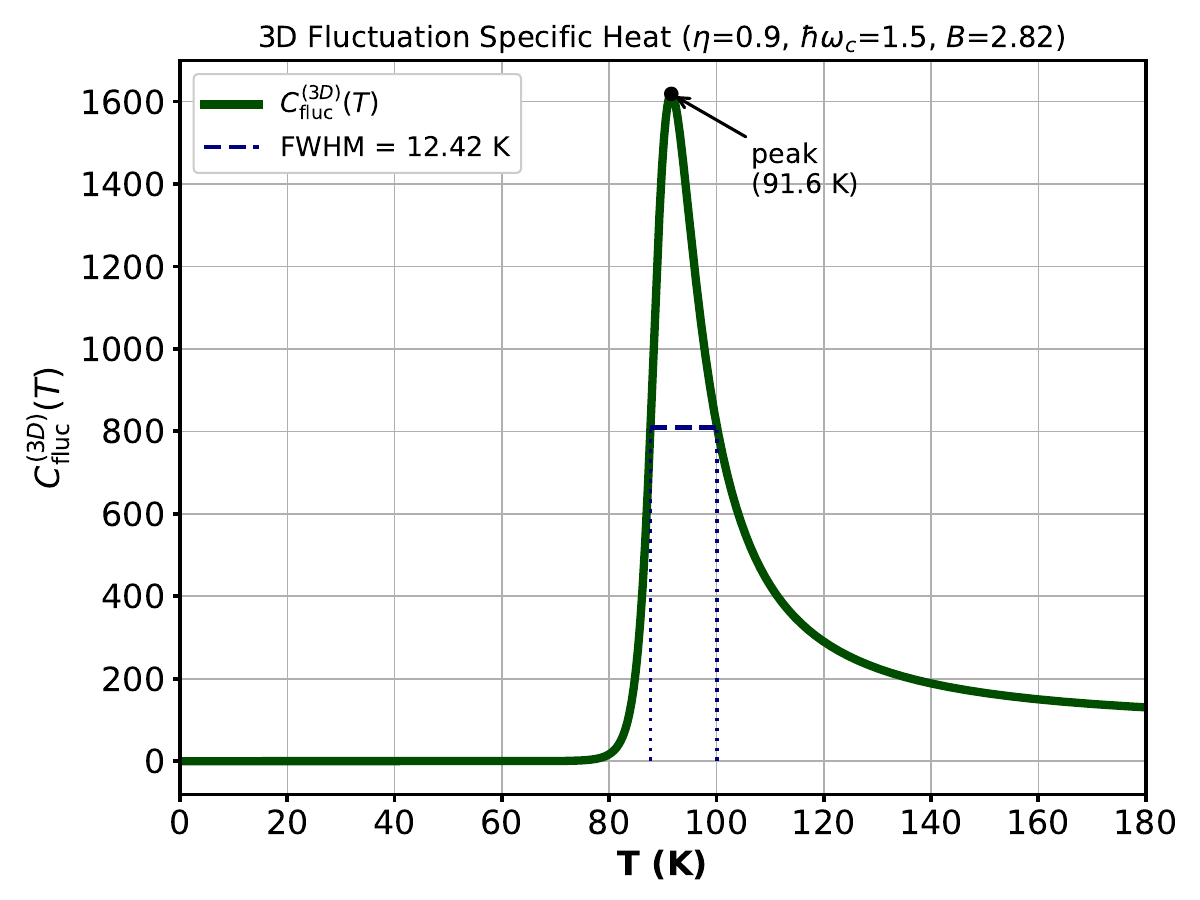}
\caption{3D fluctuation-specific heat at a fixed value of the magnetic
field $B=2.85$~T ($\hbar\omega_c=1.5$~meV) as the fluctuation parameter
$\eta$ is varied: (a) $\eta=0.3$, FWHM$=6.59$~K; (b) $\eta=0.5$,
FWHM$=8.75$~K; (c) $\eta=0.9$, FWHM$=12.42$~K. The amplitude of the
fluctuation-specific heat decreases with increasing $\eta$.}
\label{fig:8}
\end{figure}

\begin{figure}[htbp]
\centering
\includegraphics[width=0.32\linewidth]{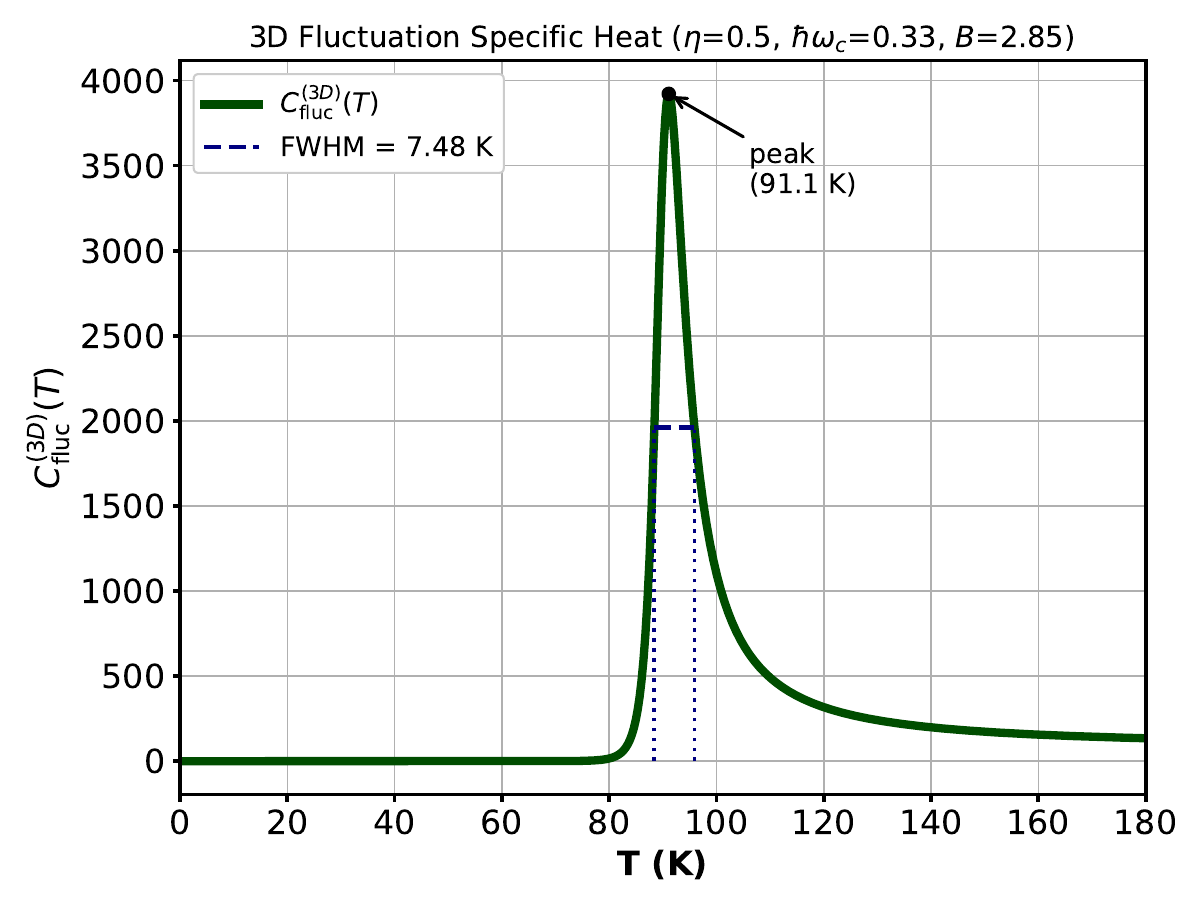}
\includegraphics[width=0.32\linewidth]{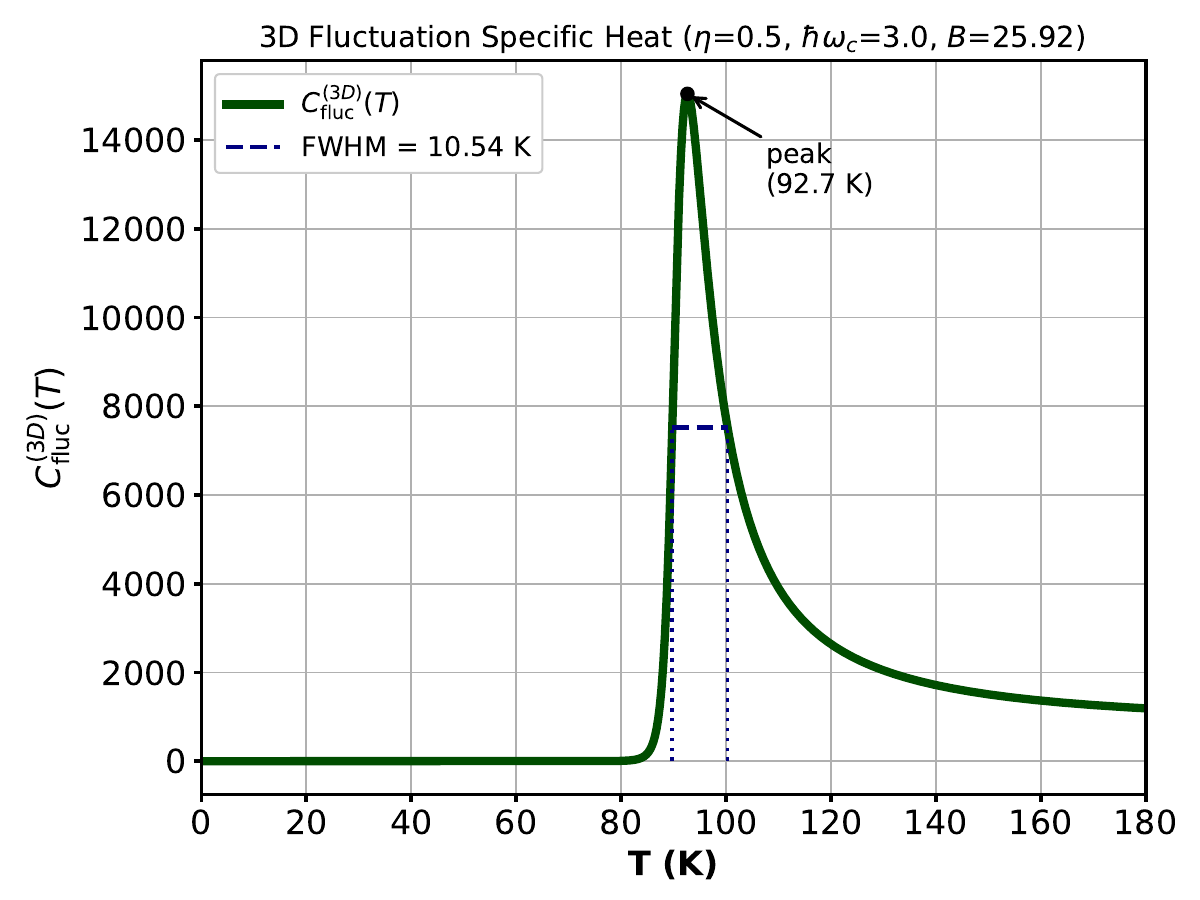}
\includegraphics[width=0.32\linewidth]{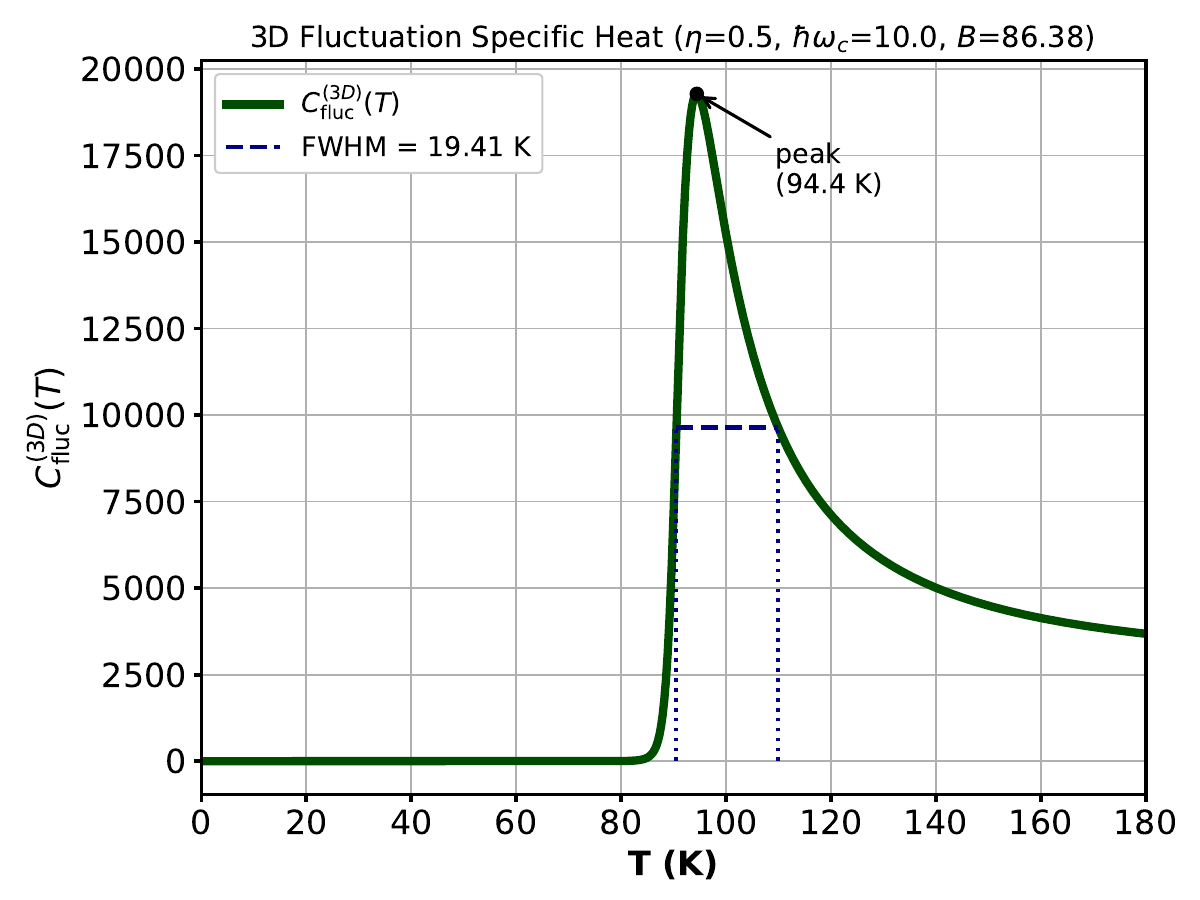}
\caption{3D fluctuation-specific heat at a fixed value of the fluctuation parameter $\eta=0.5$ as the magnetic field varies: (a) $\hbar\omega_c=0.33$~meV, $B=2.85$~T, FWHM$=7.48$~K; (b) $\hbar\omega_c=3$~meV, $B=25.92$~T, FWHM$=10.54$~K; (c) $\hbar\omega_c=10$~meV, $B=86.38$~T, FWHM$=19.41$~K. The amplitude of the fluctuation-specific heat increases with increasing magnetic field.}
\label{fig:9}
\end{figure}

At zero magnetic field, the critical transition temperature of optimally doped YBCO is $T_{c_0} \approx92$--$93$~K. The application of an external magnetic field reduces this transition temperature and significantly broadens the resistive transition, primarily due to the combined effects of vortex dynamics and anisotropy. To quantify this broadening, the fluctuation-specific heat is plotted on a grid (Figs.~\ref{fig:6}--\ref{fig:9}), and the full width at half maximum (FWHM) is extracted. The FWHM represents the temperature range over which thermal-energy fluctuations are most pronounced. A narrow FWHM suggests that ordered or cooperative fluctuations emerge within a limited temperature interval, whereas a broad FWHM, as observed with increasing $\eta$ or $B$, indicates a smeared crossover rather than a distinct transition. It should be noted that finite-size effects, impurities, or sample inhomogeneity can also cause similar broadening in experimental measurements; therefore, the FWHM obtained from theoretical analysis should be compared with experimental data with this consideration in mind.

Figure~\ref{fig:6} presents data for a fixed $B=2.85$~T and varying $\eta$ values of 0.3, 0.5, and 0.9. The same $\eta$-suppression pattern observed in Fig.~\ref{fig:1} is evident, but it is now referenced to the actual YBCO $T_{c_0} \sim92$--$93$~K, with a grid provided for the FWHM extraction. The amplitude decreases from approximately 1500 to 575 as $\eta$ increases from 0.3 to 0.9.

Figure~\ref{fig:7} presents data for a fixed $\eta=0.5$, with $\hbar\omega_c$ varying from 0.33 to 1.5~meV. The amplitude increases from approximately 210 to 760 as $B$ increases from 2.85 to 12.92~T. According to Eq.~\eqref{eq:C2Dlimit}, $C^{2D}_{\mathrm{fluc}}\propto B/[r(0,T)+\hbar\omega_c]^2$, so the amplitude's dependence on $B$ is
determined by whether the numerator (linear in $B$) or the denominator (approximately $B^2$ at high field) dominates. For the field values shown, the product of the magnetic field and the derivative $r'(0,T)$ present in the numerator appears to dominate, leading to an increasing amplitude, which aligns with the expected physical behavior.

The role of the fluctuation coupling $\eta$ is illustrated in Figs.~\ref{fig:6} and \ref{fig:8}, which isolate its effect at a fixed field in two and three dimensions, respectively. In both scenarios, increasing $\eta$ reduces the peak amplitude and broadens the anomaly. This qualitative trend, previously observed at the idealized $T_{c_0} =5$~K scale in Figs.~\ref{fig:1} and \ref{fig:3}, is now confirmed at the physically realistic YBCO transition temperature. The consistency across these distinct temperature scales supports the interpretation of $\eta$ as an intrinsic material parameter rather than an artifact of the numerical window selected for illustration.

Figure~\ref{fig:8} presents results for a fixed $B=2.85$~T and varying values of $\eta=0.3,0.5,0.9$. The consistent suppression observed with increasing $\eta$ (approximately 3400 to 2400 to 1600) further demonstrates that the regularizing effect of $\eta$ remains robust across different dimensionalities and temperature scales.

Figure~\ref{fig:9} presents data for a fixed $\eta=0.5$ and varying $\hbar\omega_c$ from 0.33 to 10~meV. The amplitude increases with increasing field (approximately $3900\to15000\to19000$).

Figures~\ref{fig:7} and \ref{fig:9} show the magnetic field's influence by isolating its effect at a fixed $\eta=0.5$ in two and three dimensions, respectively. At low to intermediate magnetic field strengths, where $\hbar\omega_c$ is comparable to or less than $r(D_{\mathrm{eff}},T)$, the numerator dominates, resulting in an increase in amplitude with $B$, as observed in Figs.~\ref{fig:2}, \ref{fig:4}, and \ref{fig:7}. At sufficiently high magnetic fields, where $\hbar\omega_c$ dominates the denominator, the amplitude decreases with $B$, as shown in Fig.~\ref{fig:8}. In strong magnetic fields, such as those present in quantum transport or low-dimensional electron systems,
the field can restrict motion along specific directions through mechanisms such as Landau quantization. Consequently, the system displays behavior indicative of reduced active dimensionality. This observed monotonicity exemplifies the ``dimensional depletion'' mechanism, which describes the reduction in effective dimensional degrees of freedom.

All these figures (Figs.~\ref{fig:6}--\ref{fig:9}) are particularly relevant because they directly correspond to the experimental setup described in Figs.~22 and 23 of Ref.~\cite{Phillips}, in which comprehensive, high-precision measurements on a single crystal with $\mathbf B\perp c$ and $\mathbf B\parallel c$ revealed broadening, lowering, and a shift of the anomaly as the field increased, depending on the field direction. We interpret these measurements in terms of fluctuations and finite-size scaling effects.

\begin{table}[htbp]
\centering
\caption{Comparison of the magnetic field's role at fixed $\eta=0.5$ in two- and three-dimensional models.}
\label{tab:2}
\begin{tabular}{ccccc}
\toprule
$\hbar\omega_c$ (meV) & $T^{2D}_{\mathrm{peak}}$ (K) & FWHM$^{2D}$ (K) &
$T^{3D}_{\mathrm{peak}}$ (K) & FWHM$^{3D}$ (K) \\
\hline
0.33  & 92.46 & 22.91 & 91.11 & 7.48  \\
1.5   & 93.70 & 22.73 & 91.99 & 8.75  \\
3.0   & 94.85 & 23.57 & 92.66 & 10.54 \\
10.0  & 98.12 & 30.27 & 94.44 & 19.41 \\
\hline
\end{tabular}
\end{table}

The magnetic field systematically shifts $T_{\mathrm{peak}}$ to higher values in both dimensions. In both scenarios, increasing $\hbar\omega_c$ causes the peak to move monotonically toward higher temperatures, progressively diverging from $T_{c_0} =92$~K. This observation directly supports the established physical mechanism: $\hbar\omega_c$ appears in the effective coefficient as $r(D,T,B)=r(D,T)+\tfrac12\hbar\omega_c$, where the cancellation of this term determines the field transition temperature, $T_{c_B}$, which is distinct from $T_{c_0}$. A stronger magnetic field shifts this resonance condition to higher temperatures, consistent with interpreting Landau levels as an additive regulating term.

Table~\ref{tab:2} shows that the shift is substantially more pronounced in two dimensions than in three. At $\hbar\omega_c=10$~meV, the 2D peak shifts by $+6.1$~K above $T_{c_0}$ (98.12~K), whereas in 3D the shift is only $+2.4$~K (94.44~K), indicating that the effect is more than twice as large in reduced dimensions. This pronounced difference arises because the transverse confinement imposed by Landau levels exerts a stronger influence on an already 3D system. In 2D, all spatial degrees of freedom are quantized by the field, so no unconstrained direction remains to mitigate the effect. In contrast, in 3D, integration over $k_z$ partially reduces the system's sensitivity to the field.

The FWHM exhibits greater sensitivity to the magnetic field in two dimensions than in three dimensions, although this effect is less pronounced than that of $\eta$. In both dimensionalities, FWHM increases with $\hbar\omega_c$; however, the quantitative impact is significantly less substantial in two dimensions. Specifically, in two dimensions,
FWHM varies from 22.9 to 30.3~K (an increase of 32\%) across the tested range of $\hbar\omega_c$, whereas in three dimensions it nearly triples, from 7.5 to 19.4~K (an increase of 160\%). These results indicate that the magnetic field is more effective at broadening the critical region in three dimensions, where it serves as a mechanism for dimensionality reduction from three to one dimension. In contrast, in two dimensions the system is already at its minimum effective dimensionality, and the magnetic field only marginally increases an already large width, which is primarily governed by $\eta$, as previously established.

A consolidated physical interpretation identifies two broadening mechanisms with hierarchical roles. This comparison establishes the previously suggested hierarchy of mechanisms. In two dimensions, the system exhibits high sensitivity to intrinsic fluctuations ($\eta$) as described by the Mermin--Wagner--Hohenberg theorem. The magnetic field
modulates this already significant width without altering its fundamental physical origin. In three dimensions, mean-field theory initially describes the system well, yielding a small intrinsic width. Here, the magnetic field becomes the dominant broadening mechanism, inducing an effective dimensional reduction ($D=3\to D_{\mathrm{eff}}=1$)
that does not occur in the absence of a field. In three dimensions, the magnetic field reveals a fluctuation regime that is otherwise inaccessible, leading to a proportionally stronger effect on the FWHM. In contrast, in two dimensions, the magnetic field amplifies an existing fluctuation regime, resulting in a more modest effect on the width but a
more pronounced temperature shift.

The dimensional character and the breakdown of standard 3D theory are central to understanding YBCO's behavior. The layered structure of YBCO renders this dimensional crossover directly relevant to its intrinsic quasi-2D electronic properties. In the purely 2D limit, thermodynamic fluctuations under an applied field are significantly enhanced due to
reduced dimensionality and a high critical temperature, thereby promoting Kosterlitz--Thouless-type vortex unbinding and Aslamazov--Larkin-type scaling while suppressing Maki--Thompson contributions. Conversely, in the 3D limit, the applied field directly suppresses $T_{c_0}$, reduces the Gaussian fluctuation regime, and can induce competition with
charge-density-wave order. In both dimensional limits, the absence of a divergence in the fluctuation-specific heat for $\eta>0$ indicates that the superconducting transition is not purely second-order or mean-field-like in the classical Ginzburg--Landau framework. Instead, this behavior reflects a breakdown of standard 3D GLT driven by low-dimensional confinement, strong thermal phase fluctuations, or competing orders. Within the present analytical framework, this breakdown is captured by the fluctuation parameter $\eta$, which introduces a continuous background of scattering-state contributions. This mechanism regularizes the critical singularity and continuously redistributes spectral weight across the transition, rather than allowing it to collapse into a single divergent point.

In underdoped cuprates, the absence of a diverging specific-heat anomaly is frequently attributed to the presence of preformed Cooper pairs at temperatures significantly above $T_{c_0}$. Unlike conventional BCS superconductors, where pairing and condensation occur simultaneously, these pairs form at higher temperatures but do not exhibit global phase
coherence. Consequently, the superconducting transition temperature is more accurately described as a phase-ordering transition governed by phase fluctuations of the OP, rather than by amplitude fluctuations. This separation between pair formation and phase coherence aligns with, and provides a physical explanation for, the rounded, non-diverging
fluctuation-specific heat observed in Figs.~\ref{fig:6}--\ref{fig:9}. It also links the current dimensionality- and field-dependent formalism to the broader phenomenology of underdoped cuprate superconductivity.

\subsection{Origin and significance of the peaks}

This subsection addresses a central conceptual aspect of the formalism presented in this paper. The peaks of $C_{\mathrm{fluc}}$ do not indicate a phase transition in the strict thermodynamic sense, as symmetry breaking does not occur in this calculation due to the implicit condition $\varphi\equiv0$. These peaks represent maxima of a fluctuation
susceptibility rather than singularities in the free energy. Specifically, $r(0,T)$ and $r(1,T)$ are, by construction, always non-negative fluctuation propagators; they characterize massive fluctuation modes and do not correspond to a mean-field OP that changes sign. Moreover, the addition of $\hbar\omega_c>0$ to a non-negative quantity further reinforces this positivity. In these calculations, $r(0,T)$ and $r(1,T)$ exhibit this behavior: they decrease toward a minimum near $T_{c_0}$ and then increase, never reaching zero, in contrast to the mean-field coefficient $r_0(T-T_{c_0})$, which becomes exactly zero at $T_{c_0}$.

The peak indicates the temperature at which the system's response to thermal fluctuations of the superconducting order parameter is maximal, representing the point where the system approaches, but does not attain, long-range order. This behavior aligns with expectations above a true transition in the regime of Gaussian fluctuations, such as
paraconductivity described by the Aslamazov--Larkin term. In this calculation, the system is sensitive to the proximity of a transition without undergoing it, since $r(D,T,B)>0$ everywhere precludes any real symmetry breaking. Within this model, the peak arises when the ratio $r'(D,T)/[r(D,T)+\hbar\omega_c]$ reaches its maximum, typically occurring
when $r(D,T)$ attains a local minimum near zero (while remaining positive due to $\eta>0$ and $\hbar\omega_c>0$), and $r'(D,T)$ (the slope) remains finite or increases at this point.

In standard, non-renormalized GLT, $C_{\mathrm{fluc}}(T)\propto1/r(T)^2$ diverges at $T_{c_0}$, where $r(T)=r_0(T-T_{c_0})=0$. The Hartree renormalization mechanism, involving $r(0,T)$ and $r(1,T)$, is specifically implemented to regularize this non-physical divergence by ensuring that $r(D,T)$ does not vanish. As a result, the infinite divergence is replaced by a finite but pronounced peak. The observed peak thus represents a softened residual signature of the underlying mean-field transition, serving as a finite ``ghost'' of the original divergence.

Although $r(D,T)$ never vanishes, it retains the structure of the mean-field equation from which it is derived (specifically, $r(D,T)\to r_0(T-T_{c_0})$ as $\eta\to0$). Consequently, the minimum of $r(D,T)$ remains closely associated with the vicinity of $T_{c_0}$. This explains why all calculated peaks consistently appear between 91 and
94~K, regardless of the values of $\eta$ or $\hbar\omega_c$ considered. The peak position is thus inherited from $T_{c_0}$ by continuity, even though the actual transition is entirely smoothed.

The peak is therefore a physically meaningful and relevant feature, as it quantifies both the intensity and width of the precursor region of fluctuations. However, it should not be mistaken for the transition itself, which is determined by the sign-change coefficients (e.g., $r(2,T)$) employed separately in vortex calculations.

\subsection{Physical significance of the shift in peak position $T_{\mathrm{peak}}$}

Across all calculated curves, the peak position shifted variably relative to $T_{c_0}=92$~K depending on the parameter being varied; in some cases, the peak remained nearly stationary, while in others it shifted by several kelvins. This shift is not a numerical artifact but has a precise physical significance, being directly related to the field-renormalized transition temperature, $T_{c_B}$, as introduced earlier in the formalism.

The maximum of $C_{\mathrm{fluc}}(T)$ is observed when the denominator $\Big[r(D,T)+\tfrac12\hbar\omega_c\Big]^p$ (where $p=2$ or $3/2$) attains its minimum value relative to $r'(D,T)^2$. This condition corresponds to the point where the field-renormalized effective quadratic coefficient, $r(D,T,B)=r(D,T)+\tfrac12\hbar\omega_c$, approaches zero. This coefficient, rather than $r(D,T)$ alone, determines the true critical condition in the presence of a magnetic field, as established previously in Eq.~\eqref{eq:Bc2}. Consequently, the observed peak serves as a direct thermodynamic signature of $T_{c_B}$, the effective transition temperature in the field, which is distinct from $T_{c_0}$, the transition temperature in zero field.

\begin{table}[htbp]
\centering
\caption{Analysis of the observed peak displacement in relation to the
swept parameter.}
\label{tab:3}
\begin{tabular}{p{2.6cm}p{5.4cm}p{6.3cm}}
\hline
Swept parameter & Peak behavior & Physical meaning \\
\hline
$\eta$ (fixed $\hbar\omega_c$) &
Remains nearly stationary, within approximately 92.4 to 92.7~K. &
Reflects the intrinsic fluctuation coupling of the material: $\eta$
broadens the transition while minimally altering its central position,
since it does not directly influence the resonance condition
$r(D,T,B)\approx0$. \\
\hline
$\hbar\omega_c$ (fixed $\eta$) &
Increases with a systematic, monotonic displacement. &
The magnetic field, through the regulating term $\tfrac12\hbar\omega_c$,
shifts the effective cancellation condition. Consequently, $T_{c_B}$ differs
from $T_{c_0}$ in direct proportion to the field strength. \\
\hline
\end{tabular}
\end{table}

In two dimensions, increasing $\hbar\omega_c$ systematically shifts the peak to higher temperatures (92.44 to 93.56~K). In three dimensions, a downward shift occurs at low $\hbar\omega_c$ (92.20 to 91.11~K when reducing $\hbar\omega_c$ from 1.5 to 0.33~meV), followed by an upward shift at higher fields (reaching 94.44~K at $\hbar\omega_c=10$~meV). This behavior highlights the asymmetry in the coupling between the field term and the dimensional term in $r(D,T)$. In three dimensions, the cubic term (free $k_z$) alters the balance between the thermal fluctuation term and the Landau quantization term, in contrast to the 2D case, where all transverse dynamics are already confined to Landau levels.

This phenomenon corresponds to what is measured in specific-heat experiments under variable magnetic fields: the position of the anomalous peak is plotted as a function of $B$, directly yielding the upper critical field curve $B_{c_2}(T)$ (or, equivalently, $T_{c_B}$). This is the same quantity measured experimentally, where a moderate shift in the
transition onset under the field was observed, in contrast to a pronounced suppression of the jump amplitude, as discussed for instance in Ref.~\cite{Phillips}.

The shift in $T_{\mathrm{peak}}$ represents a direct thermodynamic indicator of the field dependence of the critical temperature, $T_{c_B} \ne T_{c_0}$, determined by the Landau term $\tfrac12\hbar\omega_c$ in the renormalized coefficient. In contrast, the peak width, as discussed previously, is primarily determined by the fluctuation coupling $\eta$
and the effective dimensionality $D_{\mathrm{eff}}$. These two observables (position and width), therefore, encode complementary and largely independent physical information: the location of the transition under the field and the extent over which fluctuations dominate.

\subsection{Topological structure of Landau levels}

The superconducting phase of bulk superconductors is defined by the presence of long-range order (LRO). In 1972, Berezinskii \cite{Berezinskii, Rodriguez}, and independently Kosterlitz and Thouless \cite{Kosterlitz}, proposed that in 2D systems, LRO in the low-temperature phase is replaced by quasi-long-range order (quasi-LRO). Quasi-LRO means the order-parameter correlation function decays, but not exponentially, as observed above the phase-transition temperature. Therefore, while LRO does not exist at any finite temperature in 2D systems, the correlation decay transitions from a power law at low temperatures to exponential decay at high temperatures. Consequently, a phase transition is anticipated at the temperature where the correlation function changes its decay behavior.

\begin{figure}[htbp]
\centering
\includegraphics[width=0.5\linewidth]{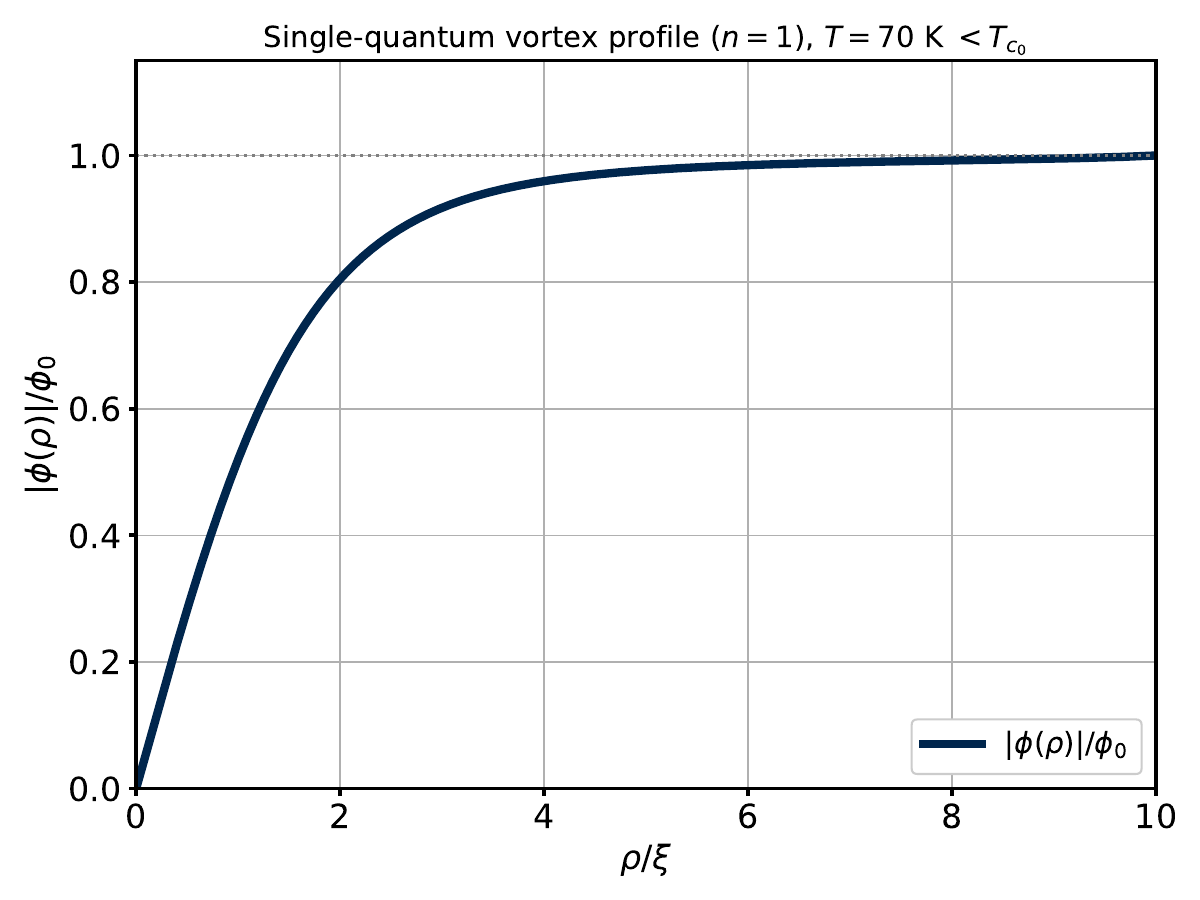}
\caption{Radial amplitude profile $|\varphi(\rho)|/\varphi_0$ of an isolated vortex ($\eta=0.5$).}
\label{fig:10}
\end{figure}

\begin{figure}[htbp]
\centering
\includegraphics[width=0.9\linewidth]{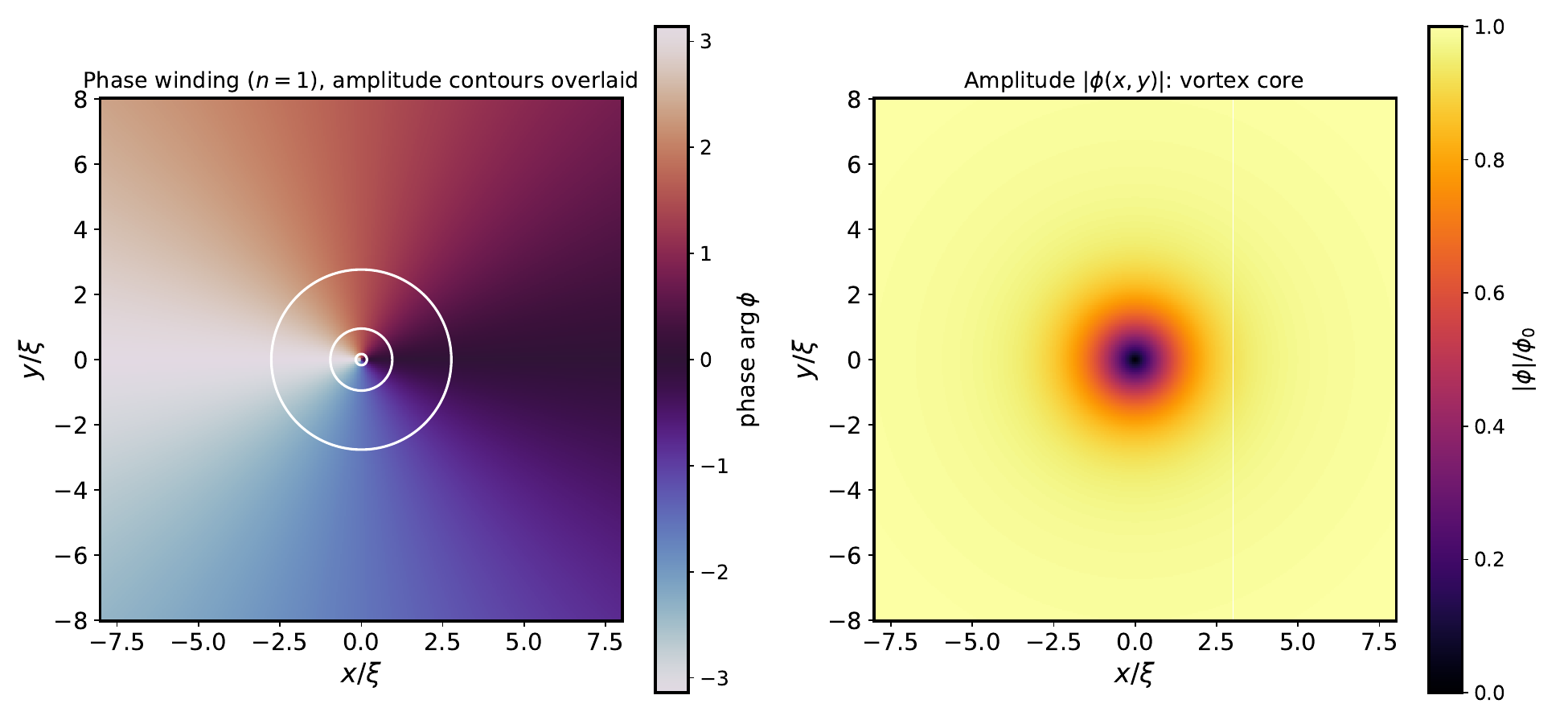}
\caption{2D phase (left) and amplitude (right) map of an isolated, singly-quantized vortex ($\eta=0.5$; winding number $n=1$).}
\label{fig:11}
\end{figure}

In previous calculations of $C_{\mathrm{fluc}}$, the Landau levels have been considered solely as an energy spectrum $E_N=\hbar\omega_c(N+\tfrac12)$ for summing a scalar thermodynamic quantity. However, the topological structure responsible for the nontrivial degeneracy is not present in the energy spectrum. Instead, it resides in the geometry of the wave functions, specifically the Berry curvature, and in the vortex structure of the superconducting state below $T_{c_0}$.

Within the framework employing a complex OP, the most direct method to explicitly solve a Ginzburg--Landau vortex and demonstrate the number of  windings is to solve the Euler--Lagrange equation for $\mathcal H_{\mathrm{GL}}[\varphi^*]$ using an axially symmetric ansatz that incorporates an explicit phase winding:
\begin{equation}
\varphi^*(\mathbf r)=f(\rho)e^{in\theta},\qquad f(0)=0,\quad f(\infty)=\varphi_0^*.
\label{eq:vortexansatz}
\end{equation}
The function $\varphi_0^*$ represents the homogeneous solution employed to calculate $\Delta C(T_c)$, the specific-heat jump. The variables $\rho$ and $\theta$ denote the polar coordinates in the $x$--$y$ plane, while $n\in\mathbb Z$ indicates the winding number, which serves as the topological invariant. The 2D transition, in the absence of an external magnetic field, is governed by the unbinding of thermally induced vortex-antivortex pairs. In this context, a vortex is characterized as a point defect, or topological defect, exhibiting zero amplitude of the OP at its center and a singularity in the phase $\theta(\mathbf r)$. The line integral along a closed, clockwise path
encircling the vortex is given by $\oint\nabla\theta\cdot d\mathbf l=2\pi n$. Utilizing the gauge relation $\mathbf A=\tfrac{\hbar c}{2e}\nabla\theta$ at distances far from the core yields
\begin{equation}
\Phi=\int\mathbf A\cdot d\mathbf l=n\Phi_0,\qquad \Phi_0=\frac{\hbar c}{2e}.
\label{eq:fluxquantum}
\end{equation}
Here, $n$ is an integer referred to as the vorticity of the vortex, also known as the topological charge or winding number. When $n=+1$, the topological defect is termed a vortex, whereas an excitation with $n=-1$ is called an antivortex \cite{Ktitorov, Herok}. Generally, the winding number  quantifies the number of times the phase changes by $\pm2\pi$ when encircling the defect in the clockwise direction. Eq.~\eqref{eq:fluxquantum} defines the superconducting flux quantum, a precise topological result
that remains independent of microscopic parameters. When a vortex is present, the line integral $\oint\nabla\theta(\mathbf r)\cdot d\mathbf l$ around a closed loop enclosing the vortex center is finite, resulting in a supercurrent circulating around the vortex core. In the absence of an external magnetic field, the supercurrent velocity $\mathbf v_s$ is
related to the phase gradient $\nabla\theta(\mathbf r)$ by $\mathbf v_s(\mathbf r)=\tfrac{\hbar}{m^*}\nabla\theta(\mathbf r)$. Consequently, phase gradients contribute to the kinetic energy density.

Dimensional characteristics and intrinsic properties of the system are analyzed by solving the Ginzburg--Landau radial equation:
\begin{equation}
\frac{\hbar^2}{2m^*}\left[-\frac1\rho\frac{d}{d\rho}\left(\rho\frac{df}{d\rho}\right)+\frac{n^2}{\rho^2}f\right]
+r(2,T)f+bf^3=0.
\label{eq:radial}
\end{equation}
This equation is typically solved numerically. The topological structure is visualized by plotting $|f(\rho)|$, which vanishes at the center and approaches $\varphi_0$ at large distances, indicating a normal core with a radius of approximately $\xi(T)$. Furthermore, plotting the phase $n\theta$ explicitly reveals the $2\pi n$ phase winding.

We numerically solve the Ginzburg--Landau radial equation for a vortex carrying one flux quantum ($n=1$), derived from $\mathcal H_{\mathrm{GL}}[\varphi^*]$ using the ansatz
$\varphi^*(\rho,\theta)=f(\rho)e^{in\theta}$. The equation is expressed in dimensionless variables ($x=\rho/\xi$, $g=f/\varphi_0^*$):
\begin{equation}
g''+g'/x-g/x^2+g-g^3=0,
\end{equation}
subject to the boundary conditions $g(0)=0$ and $g(\infty)=1$.

Unlike specific-heat curves, which utilize the branch $r(0,T)\ge 0$ suitable for normal fluctuations, a vortex solution is present only in the ordered phase where the coefficient is negative. Therefore, the closed renormalized coefficient $r(2,T)=r_0(T-T_{c_0})+\eta T/T_{c_0}$ is evaluated at $T=70$~K, below $T_{c_0} =92$~K. At this temperature, $r(2,70)=-32.77<0$, fulfilling the condition required for the existence of an ordered state that breaks $U(1)$ symmetry.

The radial amplitude profile shown in Fig.~\ref{fig:10} indicates that the normalized amplitude $|\varphi(\rho)|/\varphi_0$ vanishes at the center ($\rho=0$), increases on the $\xi$ scale, and approaches unity at large distances from the core. This behavior matches the universal form predicted by de Gennes and Tinkham \cite{Gennes, Clem}. In the 2D phase and amplitude map in Fig.~\ref{fig:11}, the left panel illustrates the explicit $2\pi$ phase winding around the core using a cyclic color map with superimposed amplitude contours. The right panel displays the characteristic density depletion at the vortex center. Integration of the phase around a closed loop encircling the vortex yields $n=1$, confirming that the winding number, as a topological invariant, is quantized independently of microscopic parameters such as mass or the quartic coefficient.

\section{Conclusion}
\label{sec:conclusion}

In this study, we developed a renormalized Ginzburg--Landau framework to describe the fluctuation-specific heat of layered high-temperature superconductors in an applied magnetic field, with YBa$_2$Cu$_3$O$_{7-\delta}$ as the representative system. The central theoretical innovation is the introduction of a dimensionality-dependent
quadratic coefficient $r(D,T)$, determined self-consistently via a Hartree-type decoupling of the $\varphi^4$-interaction. This approach replaces the conventional linear-in-temperature coefficient $r_0(T-T_{c_0})$ found in standard mean-field theory. This modification leads to two physically distinct consequences, which are addressed
jointly in this work:

\begin{itemize}
\item \textbf{Fluctuation renormalization via $\eta$.} In the absence of an applied field, retaining the $\varphi^4$ self-interaction through the material-specific coupling $\eta$ eliminates the unphysical divergence of the specific heat at $T_{c_0}$. Instead, it produces a finite, rounded maximum whose amplitude decreases and whose width broadens monotonically as $\eta$ increases. This behavior was observed consistently across all dimensionalities and temperature ranges examined (Figs.~\ref{fig:1},~\ref{fig:3},~\ref{fig:6},~\ref{fig:8}). Furthermore, the formalism recovers the sharp mean-field-like singularity in the $\eta\to0$ limit, providing an internal consistency check.
\item \textbf{Dimensional depletion via the magnetic field.} Independently of $\eta$, the application of a magnetic field quantizes the transverse degrees of freedom into Landau levels, resulting in an effective dimensional reduction, $D\to D-2$, near the upper critical field. Consequently, bulk 3D systems behave as effectively 1D, while thin-film 2D systems behave as effectively 0D. This dimensional reduction, described analytically in Eqs.~\eqref{eq:C3D}--\eqref{eq:C2Dlimit}, determines both the field dependence of the fluctuation amplitude and its behavior.
\end{itemize}

Together, fluctuation-coupling renormalization and field-induced dimensional depletion serve as complementary mechanisms that yield the same qualitative outcome: suppression and rounding of the specific-heat anomaly relative to the sharp divergence predicted by classical 3D Ginzburg--Landau theory. This framework provides a natural theoretical explanation for the long-standing experimental observation that the specific-heat jump in YBCO and related cuprates is broadened rather than divergent, without necessitating a departure from a continuous transition. By anchoring the model to YBCO's physical parameters ($T_{c_0} \approx 92$--$93$~K, coherence length $\xi_0\approx1$~nm),
we reproduce the expected sensitivity of the fluctuation region to the material's short coherence length and quasi-2D layered structure. This is consistent with the experimental observation that fluctuation effects are prominent in cuprates but negligible in conventional BCS superconductors with much larger $\xi_0$. Since the present analysis retains only the lowest ($N =0$) Landau orbital, the results are expected to be quantitatively reliable near $T_{c_B}$ and at fields approaching $B_{c2}$ across the studied temperature range. Extending the calculation to include higher Landau levels and incorporating the Zeeman (spin-paramagnetic) pair-breaking term, which was neglected in favor of the orbital contribution, represents a logical next step. This would allow for testing whether the field-driven crossover between rising- and falling-amplitude regimes persists when spin effects are included, and would enable a more quantitative fit to the FWHM data across the full $\delta$-doping range of YBa$_2$Cu$_3$O$_{7-\delta}$.


\section*{References}

\begin{thebibliography}{100}

\bibitem{Landau1} L. D. Landau and E. M. Lifshitz, {\color{blue} Fluid Mechanics}, Vol. 6 (Pergamon Press, Oxford, 1987).

\bibitem{HohenbergPC} P. C. Hohenberg and B. I. Halperin, {\color{blue} Theory of dynamic critical phenomena}, {\it Rev. Mod. Phys.} \textbf{49 (3)} (1977) 435-479.

\bibitem{Landau2} L. D. Landau, E. M. Lifshitz, and L. P. Pitaevskii, {\color{blue} Statistical Physics} Part 1, Vol. 5 (Pergamon Press, Oxford, 1994) Chap. XIV.

\bibitem{Ginzburg} V. L. Ginzburg, {\color{blue} Some remarks on phase transitions of the second kind and the microscopic theory of ferroelectric materials}, {\it Fiz. Tverd. Tela} \textbf{2 (9)} (1960) 2031-2043, [{\it Sov. Phys. Solid State} \textbf{2 (9)}, (1961) 1824-1834].

\bibitem{Stanley} H. E. Stanley, {\color{blue} Introduction to Phase Transition and Critical Phenomena}, (Oxford University Press, New York, 1971).

\bibitem{Privman} V. Privman,  P. C. Hohenberg, and A. Aharony , {\color{blue} Universal Critical-Point Amplitude Relations}, in Phase Transitions and Critical Phenomena, Vol. \textbf{14}, edited by C. Domb and J. L. Lebowitz (Academic Press, London 1991), pp. 1-134.
    
\bibitem{Larkin} A. Larkin and A. A. Varlamov, {\color{blue} Fluctuation Phenomena in Superconductors} (Clarendon Press, Oxford, 2005). 

\bibitem{Hohenberg1} P. C. Hohenberg, A. P. Krekhov {\color{blue} An introduction to the Ginzburg-Landau theory of phase transitions and nonequilibrium patterns}, {\it Phys. Rep.} \textbf{572} (2015) 4. 

\bibitem{Wu} G. Wu, Y. Xia, and S. Yang, {\color{blue} Buckling, symmetry breaking, and cavitation in periodically micro-structured hydrogel membranes}, {\it Soft Matter} \textbf{10}, (2014) 1392-1399.
       
\bibitem{Keumo3} O.C. Feulefack, C. Tsague Fotio, R.M. Keumo Tsiaze, S.E. Mkam Tchouobiap, J. E. Danga, A.J. Fotue, M.N. Hounkonnou, {\color{blue} Generation of renormalized quadratic coefficient in Landau theory: Implications for specific-heat jump calculations in high-temperature superconductors}, {\it J. Solide state Comm.} \textbf{418}   (2026) 116583  
        
\bibitem{Thouless} D. J. Thouless, {\color{blue} Critical Fluctuations of a Type-II Superconductor in a Magnetic Field}, {\it Phys. Rev. Lett.} \textbf{34} (1975) 946

\bibitem{Quader} K. F. Quader and E. Abrahams, {\color{blue} Superconducting fluctuations in specific heat in a magnetic field: Dimensional crossover}, {\it Phys. Rev B} \textbf{38} (1988) 11977. 
       
\bibitem{Gupta} B.C. Gupta 1, K.K. Nanda, {\color{blue} Specific heat of high-temperature superconductors Role of $\psi^4$ term in the Ginzburg-Landau free energy}, {\it Physica C} \textbf{265} (1996) 228-232.
        
\bibitem{Shenoy} P. A. Lee and S. R. Shenoy, {\color{blue} Effective Dimensionality Change of Fluctuations in Superconductors in a Magnetic Field},  {\it Phys. Rev Lett.} \textbf{28} (1972) 1025.   
    


\bibitem{Cybart} S. A. Cybart, E. Y. Cho, T. J. Wong, B. H. Wehlin, H. C. Ma, C. Huynh and R. C. Dynes, {\color{blue} Nano Josephson superconducting tunnel junctions in YBa$_2$Cu$_3$O$_{7 - \delta}$ directly patterned with a focused helium ion beam}, {\it Nature Nanotechnology} \textbf{10} (2015) 598.

\bibitem{Scalapino} D. J. Scalapino, M. Sears, {\color{blue}Statistical Mechanics of One-Dimensional Ginzburg-Landau Fields}, {\it Phys. Rev. B} \textbf{6} (1972) 3409.

\bibitem{Tinkham} M. Tinkham, {\color{blue} Introduction to superconductivity}, second edition, (McGraw-Hill, New York, 1996).

\bibitem{Loram} J. W. Loram, K. A. Mirza, J. R. Cooper, W. Y. Liang and J. M. Wade, {\color{blue} Electronic specific heat of YBa$_2$Cu$_3$O$_{6 +\delta}$  from 1.8 to 300 K},
{\it J. Supercond.} \textbf{7} (1994) 243.

\bibitem{Tanaka} Y. Tanaka, P.M. Shirage, A. Iyo, {\color{blue} Disappearance of Meissner Effect and Specific Heat Jump in a Multiband Superconductor, Ba$_{0.2}$K$_{0.8}$Fe$_2$As$_2$}, {\it J. Supercond Nov. Magn.}  \textbf{23} (2010) 253-256. 
     
\bibitem{Meingast} C. Meingast, A. Junod, and E. Walker, {\color{blue} Superconducting fluctuations and uniaxial-pressure dependence of $T_c$ of a Bi$_2$Sr$_2$CaCu$_2$O$_{8 + x}$ single crystal from high-resolution thermal expansion}, {\it Physica (Amsterdam) C} \textbf{272} (1996) 106.      

\bibitem{Anatoly} A. Larkin and A. A. Varlamov, {\color{blue}Fluctuation Phenomena in Superconductors} (Clerendon Press, Oxford, 2005).

\bibitem{Doniach} W. E. Lawrence and  S. Doniach, {\color{blue} In Proceedings of the 12$^{th}$ International Conference on  Low Temperature Physics}, edited by E. Kanda (Keikagu,  Tokyo, 1971), p. 361.

\bibitem{Zinn-Justin} J. Zinn-Justin, {\color{blue} Quantum Field Theory and Critical Phenomena} (Clarendon Press, Oxford, 2002).

\bibitem{Ma} S.-K. Ma, {\color{blue} Modern Theory of critical phenomena} (Benjamin, Reading, MA, 1976).

\bibitem{Amit} D. J. Amit, {\it J. Phys. C: Solid State Phys.} \textbf{7}, 3369 (1974).

\bibitem{Papon} P. Papon, J. Leblond and P.H.E. Meijer, {\color{blue} The Physics of Phase Transitions: Concepts and Applications} (Springer-Verlag, Berlin Heidelberg, 2006).

\bibitem{Varlamov} A. A. Varlamov, G. Balestrino, E. Milani, and D.V. Livanov, {\color{blue} The Role of Density of States Fluctuations in the Normal State Properties of High $T_c$ Superconductors}, {\it Adv. Phys.} \textbf{48}, p. 655 (1999).
    
\bibitem{Kleinert} H. Kleinert, and V. Schulte-Frohlinde, {\color{blue} Critical Properties of $\phi^4$-Theories}, (World Scientific, Singapore, 2001).

\bibitem{Poole} C. P. Poole, Jr., {\color{blue} Handbook of Superconductivity} (Academic  Press,  New-York, 2000).

\bibitem{Keumo1} R. M. Keumo Tsiaze, S. E. Mkam Tchouobiap, J. E. Danga, S. Domngang, M. N. Hounkonnou, {\color{blue} Renormalized Gaussian approach to critical fluctuations in the Landau-Ginzburg-Wilson model and finite-size scaling}, {\it J. Phys A: Math. Theor.} \textbf{44} (2011) 285002. 

\bibitem{Keumo2} R. M. Keumo Tsiaze, A. V. Wirngo, S. E. Mkam Tchouobiap, A. J. Fotue, E. Baloitcha, and M. N. Hounkonnou
{\color{blue}  Effects of critical fluctuations and dimensionality on the jump in specific heat at the superconducting transition temperature: Application to YBa$_2$Cu$_3$O$_{7 - \delta}$, Bi$_2$Sr$_2$CaCu$_2$O$_{8 + \delta}$, and KOs$_2$O$_6$ compounds}, {\it Phys. Rev. E} \textbf{93} (2016) 062105.  

\bibitem{Mermin} N. D. Mermin, H. Wagner, {\color{blue} Absence of Ferromagnetism or Antiferromagnetism in One- or Two-Dimensional Isotropic Heisenberg Models}, {\it Phys. Rev. Lett.} \textbf{17} (1966) 1133.

\bibitem{Hohenberg} P. C. Hohenberg, {\color{blue} Existence of Long-Range Order in One and Two Dimensions}, {\it Phys. Rev.} \textbf{158} (1967) 383.

\bibitem{Patashinski} A. Z. Patashinski and V. L. Pokrovskii,  {\color{blue} Behavior of an ordering system near the phase transition point}, {\it Soviet Physics JETP.} \textbf{23} (1966) 292.
 
\bibitem{Cooper} J. W. Loram, J. R. Cooper, J. M. Wheatley, K. A. Mirza, and R. S. Liu, {\color{blue} Critical and Gaussian fluctuation effects in the specific heat and conductivity of high-$T_c$, superconductors}, {\it Philos. Mag. B} 65 (1992) 1405.       

\bibitem{DFisher} D. S. Fisher, M. P. A. Fisher, and D. A. Huse, {\color{blue} Thermal fluctuations, quenched disorder, phase transitions, and transport in type-II superconductors}, {\it Phys. Rev. B} 43, 130 (1991).
   
\bibitem{Phillips} N. E. Phillips R.A. Fisher, J.E. Gordon, in {\color{blue} Progress in Low Temperature Physics}, edited by D. F. Brewer Amsterdam (Elsevier Science Publishers B. V., Amsterdam, 1992), Vol. 13, pp. 267-357.  
    
\bibitem{Berezinskii}  V. L. Berezinskii,  {\color{blue} Destruction of long-range order in one-dimensional and two-dimensional systems possessing a continuous symmetry group. II. Quantum systems}, {\it Zh. Eksp. Teor. Fiz.} \textbf{61} (1972) 1144  [{\it Sov.Phys. JETP} \textbf{34}  (1972)  610].

\bibitem{Rodriguez} J. P. Rodriguez, {\color{blue} Berezinskii-Kosterlitz-Thouless transition in a spin-charge-separated superconductor},  {\it Phys. Rev B} \textbf{49} (1994) 9831.

\bibitem{Kosterlitz}  J. M. Kosterlitz and D.J. Thouless, {\color{blue} Ordering, metastability and phase transitions in two-dimensional systems},   {\it J. Phys. C} 6 (1973) 1181.      
    
\bibitem{Ktitorov} S. A. Ktitorov, {\color{blue} Fractal vortex structure in a superconductor lattice model}, {\it Tech. Phys. Lett.} 29 (2003) 181-183.  

\bibitem{Herok} R. Szczesniak, A. P. Durajski, and L. Herok, {\color{blue} Theoretical description of the SrPt superconductor in the strong-coupling limit}, {\it Phys. Scr.} \textbf{89} (2014) 125701.     

\bibitem{Gennes} P. G. De Gennes, M. Tinkham,  {\color{blue} Magnetic Behavior of Very Small Superconducting Particles}, {\it Physics Physique Fizika} \textbf{1} (1964) 107. 

\bibitem{Clem} J. R, Clem,  {\color{blue} Simple model for the vortex core in a type-II superconductor},  {\it Phys. Rev B} \textbf{12} (1975) 174-178

\end{thebibliography}
\end{document}